\documentclass[preprint,12pt]{elsarticle}

\usepackage{geometry}
\usepackage{amsmath}
\usepackage{esint}
\usepackage{algorithm2e}

\usepackage{lineno,hyperref}
\hypersetup{colorlinks=true}

\journal{Journal of Computational Physics}

\newcommand{\bnabla}{\boldsymbol{\nabla}}
\newcommand{\bcdot}{\boldsymbol{\cdot}}
\newcommand{\p}{\partial}
\newcommand{\ffrac}[2]{\ensuremath{\frac{\displaystyle #1}{\displaystyle #2}}}

\biboptions{sort&compress}

\begin{document}

\begin{frontmatter}

\title{A multigrid solver for the coupled pressure-temperature equations in an all-Mach solver with VoF}

\author[add1]{Youssef Saade\corref{correspondingauthor}}
\cortext[correspondingauthor]{Corresponding author}
\ead{y.saade@utwente.nl}

\author[add1,add2]{Detlef Lohse}

\author[add3]{Daniel Fuster}

\address[add1]{Physics of Fluids Group, Max Planck Centre for Complex Fluid Dynamics and J.M. Burgers Centre for Fluid Dynamics, University of Twente, P.O. Box 217, 7500 AE Enschede, The Netherlands}
\address[add2]{Max Planck Institute for Dynamics and Self-Organisation, Am Fassberg 17, 37077 Göttingen, Germany}
\address[add3]{Sorbonne Université, Centre National de la Recherche Scientifique, UMR 7190, Institut Jean Le Rond $\p$’Alembert, F-75005 Paris, France}

\begin{abstract}
	We present a generalisation of the all-Mach solver of Fuster \& Popinet (2018) \cite{jcp-fuster-2018} to account for heat diffusion between two different compressible phases. By solving a two-way coupled system of equations for pressure and temperature, the current code is shown to increase the robustness and accuracy of the solver with respect to classical explicit discretization schemes. Different test cases are proposed to validate the implementation of the thermal effects: an Epstein-Plesset like problem for temperature is shown to compare well with a spectral method solution. The code also reproduces free small amplitude oscillations of a spherical bubble where analytical solutions capturing the transition between isothermal and adiabatic regimes are available. We show results of a single sonoluminescent bubble (SBSL) in standing waves, where the result of the DNS is compared with that of other methods in the literature. Moreover, the Rayleigh collapse problem is studied in order to evaluate the importance of thermal effects on the peak pressures reached during the collapse of spherical bubbles. Finally, the collapse of a bubble near a rigid boundary is studied reporting the change of heat flux as a function of the stand-off distance.
\end{abstract}


\end{frontmatter}

\section{Introduction}\label{sec::1}
Bubble cavitation is relevant in many engineering processes. The inception of a bubble and its interaction with nearby boundaries is sometimes intended, sometimes not. Examples of the former scenario are: laser-induced forward transfer (LIFT) \cite{amt-serra-2019,jfm-jalaal-2019,apl-jalaal-2019,jfm-saade-2021}, lithotripsy \cite{jasa-maeda-2016}, needle-free injection technologies \cite{jap-og-2019,addr-schoppink-2022}, sonochemistry \cite{science-suslick-1990} etc. An example of the latter scenario is when bubbles nucleate in the low pressure regions behind rotating ship turbines upon which they collapse, thus inflicting damage and causing erosion \cite{arfm-blake-1987}. Understanding the complex bubble dynamics is therefore crucial for tuning and controlling these processes. Further examples are given in the review by Lohse (2018) \cite{prf-lohse-2018}. Experiments are usually a great tool of unravelling the intricate physics of bubble motion; however, there are many technical limitations that render the access to all the fluid properties quite impossible. The need thus emerges for decent numerical tools that correctly model two-phase compressible flows.

Baer  \& Nunziato (1986) \cite{ijmf-baer-1986} proposed a seven equations model where they solve for the conservation of mass, momentum and total energy in each of the two phases, as well as an equation for the volume fraction. Several authors later adapted this ``parent model" and used it to solve interface problems as well as fluid mixtures with several velocities \cite{jcp-saurel-1999,jfm-saurel-2001,jcp-zein-2010,ijmf-ha-2015}. Using an asymptotic analysis in the limit of stiff velocity relaxation only, or in the limit of both stiff velocity and pressure relaxation, one obtains a six \cite{pof-kapila-2001,jcp-saurel-2009,jcp-pelanti-2014} or five \cite{pof-kapila-2001,jcp-murrone-2005} equations model, respectively. Heat and mass transfer are taken into account through empirical relations depending on the type of process that is studied. In such methods, due to the treatment of the convective terms, pressure oscillations as well as temperature spikes occurred at discontinuities or interfaces where a jump in material properties usually exists. Johnsen \& Ham (2012) \cite{jcp-johnsen-2012} proposed approaches to overcome these spurious errors; however, temperature undershoots could still happen, thus affecting the solution when heat diffusion between the phases is taken into account. Beig \& Johnsen (2015) \cite{jcp-beig-2015} later developed an efficient treatment of the temperature, yielding better solutions in the cases of coupling via heat diffusion. Such solvers have been used to study the temperatures produced by inertially collapsing bubbles near rigid surfaces \cite{jfm-beig-2018}.

Another family of numerical methods, capable of capturing the compressibility effects, stem from the generalisation of numerical schemes developed for incompressible flows \cite{jcp-xiao-2004,jcp-xiao-2006,jcp-kwatra-2009}. In particular, the all-Mach method is appealing for the simulation of different kinds of flows ranging from subsonic to supersonic conditions. The main advantage of this method is its capability to retrieve the incompressible limit without the classical time step restriction of compressible solvers, where one needs to computes an acoustic CFL condition based on the speed of sound in the least compressible fluid. Fuster \& Popinet (2018) \cite{jcp-fuster-2018} recently proposed an all-Mach solver using the Volume-of-Fluid (VoF) method for the tracking of the interface, taking into account viscous and capillary forces. The solver, implemented in the free software program {\fontfamily{qzc}\selectfont Basilisk} \cite{jcp-popinet-2015}, previously assumed adiabatic processes only. In the present work, we extend it by taking into account heat diffusion between the phases. For this end we derive a two-way coupled system of equations for pressure and temperature which we implicitly solve using a multigrid solver. A plethora of additional phenomena where temperature is important could then be investigated.

The manuscript is organised as follows: in section \ref{sec::2}, we present the governing equations embedded in the all-Mach solver, including the newly derived two-way coupled systems of equations for pressure and temperature. In section \ref{sec::3}, we describe the employed numerical scheme, as well as the multigrid solver used for the solution of the aforementioned system of equations. In section \ref{sec::4}, we propose new test cases to validate the correct implementation of the thermal effects, ranging from the linear to the strongly non-linear regimes. In section \ref{sec::5}, we present numerical examples of the spherical and axisymmetric Rayleigh collapse of a bubble. Finally, we draw our conclusion and provide an outlook for future work.

\section{Governing equations}\label{sec::2}
The equations, governing the compressible two-phase flows considered in the present work, are presented in this section. The mass and momentum conservation equations, written in their conservative form, read,
\begin{equation}
	\frac{\p\rho_i}{\partial t} + \bnabla\bcdot\left(\rho_i\boldsymbol{u}_i\right) = 0,
	\label{eq::1}
\end{equation}
\begin{equation}
	\frac{\p\left(\rho_i\boldsymbol{u}_i\right)}{\p t} + \bnabla\bcdot\left(\rho_i\boldsymbol{u}_i\boldsymbol{u}_i\right) = \bnabla\bcdot\boldsymbol{\tau}_i,
	\label{eq::2}
\end{equation}
where the subscript $i$ denotes either phases, set throughout this paper to $1$ or $2$ for the liquid and gas phases, respectively. In equations \ref{eq::1} and \ref{eq::2}, $\rho$ is the density, $\boldsymbol{u}$ is the velocity field, $\boldsymbol{\tau}$ is Cauchy's stress tensor defined as,
\begin{equation}
	\boldsymbol{\tau}_i = -p_i\boldsymbol{I} + \mu_i\left(\bnabla\boldsymbol{u}_i + \bnabla\boldsymbol{u}_i^T\right),
	\label{eq::3}
\end{equation}
$p$ is the pressure field, $\mu$ is the dynamic viscosity, and $\boldsymbol{I}$ is the identity tensor. 
Note that, following Stokes' hypothesis \cite{tcps-stokes-1845}, the bulk viscosity is neglected. This assumption, commonly used in the analysis of compressible flows,  states that an isotropic expansion of a liquid element does not induce viscous stresses \cite{am-burseti-2015}.

In the absence of mass transfer, the velocity field is continuous across the interface $\boldsymbol{u}_1\bcdot\boldsymbol{n} = \boldsymbol{u}_2\bcdot\boldsymbol{n}$, where $\boldsymbol{n}$ is the unit  vector normal to this interface. It is therefore convenient to work in a one-fluid formulation, similarly to classical incompressible formulations, with a single continuous averaged velocity field $\overline{\boldsymbol{u}}$. Here and in the following, the bar characterizes an averaging process. The Laplace equation gives the pressure jump at the interface,
\begin{equation}
  p_1- p_2 = \sigma\kappa + \mu_1 \boldsymbol{n} \bcdot \boldsymbol{\tau}_1 \bcdot \boldsymbol{n} - \mu_2 \boldsymbol{n} \bcdot \boldsymbol{\tau}_2 \bcdot \boldsymbol{n} ,
\end{equation}
where $\sigma$ is the surface tension coefficient and $\kappa$ is the local curvature. Applying this jump condition, we obtain the averaged momentum equation which we actually solve in the one-fluid formulation,
\begin{equation}
	\frac{\p\overline{\rho\boldsymbol{u}}}{\p t} + \bnabla\bcdot\left(\overline{\rho\boldsymbol{u}}~\overline{\boldsymbol{u}}\right) = \bnabla\bcdot\overline{\boldsymbol{\tau}} + \sigma\kappa\delta_s\boldsymbol{n},
\end{equation}
where $\delta_s$ is a characteristic function only defined at the interface.

In the absence of mass transfer effects, the conservation of total energy is written as,
\begin{equation}
	\frac{\p}{\p t}\left[\rho_i\left(e_i + \frac{1}{2}\boldsymbol{u}_i^2\right)\right] + \bnabla\bcdot\left[\rho_i\left(e_i + \frac{1}{2}\boldsymbol{u}_i^2\right)\boldsymbol{u}_i\right] = \bnabla\bcdot\left(\boldsymbol{\tau}_i\bcdot\boldsymbol{u}_i\right) - \bnabla\bcdot\boldsymbol{q}_i,
	\label{eq::4}
\end{equation}
where $e$ is the specific internal energy, and $\boldsymbol{q}$ is the heat flux given by Fourier's law,
\begin{equation}
	\boldsymbol{q}_i = -k_i\bnabla T_i,
	\label{eq::5}
\end{equation}
where $k$ is the thermal conductivity, and $T$ is the temperature field. The present work takes into account heat diffusion, whereas the previous version of the all-Mach solver only considered adiabatic processes \cite{jcp-fuster-2018}. The all-Mach solver is a density based solver with the density, momentum and total energy as its primitive variables. However, in order to compute fluxes, the solver makes use of an evolution equation for pressure, similar to the Poisson equation in incompressible solvers, when projecting the velocity field to make it divergence free \cite{jcp-popinet-2003}. To obtain such an equation in the current formulation, we first write the enthalpy equation,
\begin{equation}
	\rho_i c_{p,i}\frac{DT_i}{Dt} = \beta_i T_i \frac{Dp_i}{Dt} - \bnabla\bcdot\boldsymbol{q}_i,
	\label{eq::6}
\end{equation}
where $c_p$ is the specific heat capacity, $\beta$ is the thermal dilation coefficient, and $D/Dt$ is the total derivative. We then express the density differential as the sum of both isothermal and isobaric processes,
\begin{equation}
	d\rho = \left(\frac{\p\rho}{\p p}\right)_T dp + \left(\frac{\p\rho}{\p T}\right)_p dT = \frac{\gamma}{c^2}dp - \rho\beta dT,
	\label{eq::7}
\end{equation}
using the definitions of the speed of sound $c$, the ratio of specific heats $\gamma$, and the thermal dilation $\beta$. If we then combine equations \ref{eq::1}, \ref{eq::6}, and \ref{eq::7}, we obtain an equation for pressure that reads
\begin{equation}
	\left(\frac{\gamma_i}{\rho_i c_i^2} - \frac{\beta_i^2 T_i}{\rho_i c_{p,i}}\right)\frac{Dp_i}{Dt} = -\frac{\beta_i}{\rho_i c_{p,i}}\bnabla\bcdot\boldsymbol{q}_i - \bnabla\bcdot\boldsymbol{u}_i.
	\label{eq::8}
\end{equation}
We thus have a two-way coupled system of equations \ref{eq::6} and \ref{eq::8}, for both the temperature and the pressure fields. This is to be contrasted with the previous version of the all-Mach solver, where only equation \ref{eq::8} is solved while neglecting heat transfer ($\boldsymbol{q} = \boldsymbol{0}$). To close the system of equations, an equation of state (EoS), relating the thermodynamic quantities \{$p$,$\rho$,$T$\}, is needed. We employ the Noble-Abel Stiffened-Gas (NASG) EoS which shows a better agreement with experiments than the Stiffened-Gas (SG) EoS, regarding the relation between the liquid's specific volume $v_1$ and its temperature \cite{pof-saurel-2016}, and thus a better relation between $\rho_1$ and $T_1$. The NASG EoS reads
\begin{equation}
	\rho_i e_i = \frac{p_i + \Gamma_i\Pi_i}{\Gamma_i - 1}\left(1 - \rho_i b_i\right) + \rho_i q_i,
	\label{eq::9}
\end{equation}
where $\Gamma$, $\Pi$, $b$, and $q$ are fitting parameters, different for each fluid. The values of these parameters, and other thermodynamic properties, are presented in table \ref{tab::1} for liquid water. Throughout this paper, gases are considered to be ideal, thus obeying the perfect gas EoS. The latter is retrieved from the NASG EoS by setting $\Gamma_2 = \gamma_2$, and $\Pi_2 = b_2 = q_2 = 0$.
\begin{table}
	\centering
	\begin{tabular}{c  c}
		\hline\hline
		$\Gamma_1$ & $1.19$ \\
		$\Pi_1 \ (Pa)$ & $7028 \times 10^5$ \\
		$b_1 \ (m^3/kg)$ & $6.61 \times 10^{-4}$ \\
		$q_1 \ (J/kg)$ & $ -1177788$ \\
		$c_{p,1} \ (J/kg/K)$ & $4285$ \\
		$c_{v,1} \ (J/kg/K)$ & $3610$ \\
		\hline
	\end{tabular}
\caption{NASG EoS parameters and thermodynamic properties for liquid water in the temperature range $[300:500]\ K$.}
\label{tab::1}
\end{table}
The expression of the thermal dilation coefficient, derived in the framework of NASG, is written as,
\begin{equation}
	\beta_i = \frac{1}{v_i}\left(\frac{\p v_i}{\p T_i}\right)_p = - \frac{1}{\rho_i}\left(\frac{\p \rho_i}{\p T_i}\right)_p = \frac{(\Gamma_i - 1)c_{v,i}}{(\Gamma_i - 1)c_{v,i}T_i + b_i(p_i + \Pi_i)}.
	\label{eq::10}
\end{equation}
In this EoS, the speed of sound $c$ is expressed as follows,
\begin{equation}
	c_i^2 = \frac{\Gamma_i(p_i + \Pi_i)}{\rho_i(1 - \rho_i b_i)}.
	\label{eq::11}
\end{equation}
 
 Finally, let the interface be represented by a Heaviside function $\mathcal{H}$ equal to 1 in the reference phase. The position of the interface is then tracked by solving an advection equation for $\mathcal{H}$,
 \begin{equation}
 	\frac{\p \mathcal{H}}{\p t} + \overline{\boldsymbol{u}}\bcdot\bnabla\mathcal{H} = 0.
 \end{equation}

\section{Numerical scheme}\label{sec::3}
In this section we present an overview of the numerical method, detailed in \cite{jcp-fuster-2018}, with the added steps regarding the implementation of the heat diffusion effects, and a description of the employed multigrid solver. This solver uses the volume fraction of a reference phase, the individual component of both density and total energy, as well as the averaged momentum as primitive variables. The discretization in time of the mass, averaged momentum, total energy, and volume fraction evolution equations gives a system of equations of the form,
\begin{equation}
	\frac{\boldsymbol{Y}^{n+1} - \boldsymbol{Y}^{n}}{\Delta t} + \bnabla \bcdot \boldsymbol{F}^{(adv)} = \bnabla \bcdot \boldsymbol{F}^{(non-adv)} + \boldsymbol{S},
\end{equation}
with
$$
\boldsymbol{Y} = \begin{pmatrix}
	C_1\\
	C_1 \rho_1 \\
	C_2 \rho_2 \\
	\overline{\rho \boldsymbol{u}} \\
	C_1 \rho_1 e_{T,1}\\
	C_2 \rho_2 e_{T,2}
\end{pmatrix}, 
~~~\boldsymbol{F}^{(adv)} = 
\begin{pmatrix}
	C_1 \overline{\boldsymbol{u}}\\
	C_1 \rho_1\overline{\boldsymbol{u}}\\
	C_2 \rho_2\overline{\boldsymbol{u}}\\
	\overline{\rho\boldsymbol{u}}~\overline{\boldsymbol{u}} \\
	C_1 \rho_1 e_{T,1} \overline{\boldsymbol{u}}\\
	C_2 \rho_2 e_{T,2} \overline{\boldsymbol{u}}
\end{pmatrix},
~~~\boldsymbol{F}^{(non-adv)} = 
\begin{pmatrix}
	0\\
	0\\
	0\\
	\overline{\tau} \\
	C_1 (\boldsymbol{\tau}_1 \bcdot\overline{\boldsymbol{u}} - \boldsymbol{q}_1)\\
	C_2 (\boldsymbol{\tau}_2 \bcdot\overline{\boldsymbol{u}} - \boldsymbol{q}_2)
\end{pmatrix},
~~~\boldsymbol{S} = 
\begin{pmatrix}
	C_1\bnabla\bcdot\overline{\boldsymbol{u}}\\
	0\\
	0\\
	\sigma\kappa\nabla C \\
	0 \\
	0
\end{pmatrix},
$$
where $e_{T,i} = e_i + \frac{1}{2} |\overline{\boldsymbol{u}} |^2$ is the total energy per unit mass and the colour function
$C_1$ is equal to the volume fraction of the reference phase in a control volume represented by the grid cell, and $C_2 = 1-C_1$. Note that in the absence of mass transfer, the interface is advected with the flow at a velocity $\boldsymbol{u}_s = \boldsymbol{u}_1\bcdot\boldsymbol{n} = \boldsymbol{u}_2\bcdot\boldsymbol{n}$. It is therefore convenient to work in a one-fluid formulation, similarly to classical incompressible formulations, given the continuity of the velocity field across the interface. The average of a quantity $\phi$ is defined as $\overline{\phi} = C\phi_1 + (1 - C)\phi_2$. We would then have averaged properties, such as density $\overline{\rho}$ and viscosity $\overline{\mu}$, and an averaged momentum field $\overline{\rho\boldsymbol{u}}$, out of which the flow velocity $\overline{\boldsymbol{u}}$ can be derived and used for the advection of the colour function and the conserved quantities.

An important property of the all-Mach solver is that, similar to a Riemann solver, it is based on the computation of fluxes and therefore exactly conserves mass, momentum and energy in the absence of surface tension forces. In particular, the advection fluxes $\boldsymbol{F}^{(adv)}$ are obtained by using a consistent scheme for the advection of the conserved quantities (density, momentum, and total energy) and the volume fraction $C$, as described in \cite{jcp-fuster-2018,cf-arrufat-2021}. This avoids any numerical diffusion of the conserved quantities when computing the advection fluxes, especially for high density ratios \cite{cf-arrufat-2021}. In other words, the discontinuity in these quantities is advected exactly at the same velocity as that of the moving interface. The employed VoF method is geometric, in which the interface has a sharp representation \cite{arfm-scardovelli-1999}. The numerical scheme used for the advection, proposed by Weymouth \& Yue (2010) \cite{jcp-weymouth&yue-2010}, conserves mass to machine accuracy in the incompressible limit. It is preceded by a PLIC reconstruction of the sharp interface, where the normal $\boldsymbol{n}$ to each interfacial segment is computed using the Mixed Youngs-Centered (MYC) method described in \cite{bookDNS}. The advection fluxes are computed in a directionally split manner as detailed in \cite{jcp-weymouth&yue-2010}. Without lack of generality, after this step we compute the advected values of the primitive variables,
\begin{equation}
  \label{eq:Yadv}
	\boldsymbol{Y}^{(adv)} \equiv  \boldsymbol{Y}^{n} - \Delta t \bnabla \bcdot \boldsymbol{F}^{(adv)} 
\end{equation}
as well as the updated values of the volume fraction. A predicted value of the velocity field at the end of the time-step ($\boldsymbol{u}_{pred}^{n+1}$), which already accounts for viscous and surface tension effects is obtained by implicitly solving the averaged momentum equation,
\begin{equation}
  \frac{\overline{\rho} \overline{\boldsymbol{u}}_{pred}^{n+1} - \overline{\rho \boldsymbol{u}}^{(adv)}}{\Delta t} =
  -\nabla p^n + \bnabla\bcdot\left[\mu\left(\bnabla\overline{\boldsymbol{u}}_{pred} + \bnabla\overline{\boldsymbol{u}}^T_{pred}\right)\right]^{n+1} + \sigma\kappa\nabla C^{n+1},
	\label{eq::2num}
\end{equation}
where the pressure gradient is evaluated at the previous time step.

The resulting estimation of the velocity field is finally corrected using
\begin{equation}
  \overline{\boldsymbol{u}}^{n+1} = \overline{\boldsymbol{u}}^* - \frac{\Delta t}{\overline{\rho}^{n+1}} \nabla p^{n+1},
  \label{eq::unp1}
\end{equation}
where $\overline{\boldsymbol{u}}^*$ is defined as
\begin{equation}
  \overline{\boldsymbol{u}}^* \equiv \overline{\boldsymbol{u}}^{n+1}_{pred} + \frac{\Delta t}{\overline{\rho}^{n+1}} \nabla p^n.
  \label{eq::provel}
\end{equation}
To compute the pressure gradient, we take the divergence of equation \ref{eq::unp1}
\begin{equation}
	\bnabla\bcdot\overline{\boldsymbol{u}}^{n+1} = \bnabla\bcdot\overline{\boldsymbol{u}}^* - \bnabla\bcdot\left(\frac{\Delta t}{\overline{\rho}^{n+1}}\bnabla p^{n+1}\right).
	\label{eq::17}
\end{equation}
For incompressible flows $\bnabla\bcdot\boldsymbol{u}^{n+1} = 0$ and equation \ref{eq::17} becomes a Poisson equation that is sufficient to compute the pressure field $p^{n+1}$ and the divergence-free velocity field $\boldsymbol{u}^{n+1}$ from equation \ref{eq::unp1}. Naturally, this is not the case in a compressible framework, where $\bnabla\bcdot\boldsymbol{u}^{n+1} $ is not necessarily zero, therefore leaving us with two unknowns: $\boldsymbol{u}^{n+1}$ and $p^{n+1}$, and with the need of an additional equation to close the system. In the previous version of the all-Mach solver, equation \ref{eq::8} served this purpose, neglecting, however, heat diffusion. The novelty in the present work is the implementation of the thermal effects by solving the two-way coupled system of equations \ref{eq::6} and \ref{eq::8},
\begin{equation}
  \rho c_p\frac{T^{n+1} - T^{(adv)}}{\Delta t} = \beta T^{(adv)}\frac{p^{n+1} - p^{(adv)}}{\Delta t} + \bnabla\bcdot\left(k\bnabla T^{n+1}\right),
	\label{eq::18}
\end{equation}
\begin{equation}
  \left(\frac{\gamma}{\rho c^2} - \frac{\beta^2 T^{(adv)}}{\rho c_p}\right)\frac{p^{n+1} - p^{(adv)}}{\Delta t} = \frac{\beta}{\rho c_p}\bnabla\bcdot\left(k\bnabla T^{n+1}\right) - \bnabla\bcdot\boldsymbol{u}^{*} + \bnabla\bcdot\left(\frac{\Delta t}{\rho}\bnabla p^{n+1}\right),
	\label{eq::19}
\end{equation}
where equation \ref{eq::19} is obtained by combining \ref{eq::8} and \ref{eq::17}. Note that $T^{(adv)}$ and $p^{(adv)}$ are provisional values obtained after the advection step although
not advected \textit{per se}. These fields are therefore not cloned as tracers and associated with the colour function $C$, as is the case for the conserved quantities. Rather, $p^{(adv)}$ is computed from the advected total energy via the equation of state, similarly to what has been done in \cite{jcp-xiao-2006} and \cite{jcp-jemison-2014},
\begin{equation}
  p^{(adv)} = \frac{\overline{\rho e_T}^{(adv)} - \ffrac{1}{2}\rho|\boldsymbol{u}|^2 - \overline{\ffrac{\Gamma\Pi}{\Gamma - 1}\left(1 - \rho b\right) + \rho q}}{\overline{\ffrac{1 - \rho b}{\Gamma - 1}}},
	\label{eq::20}
\end{equation}
where $\overline{\rho e_T}^{(adv)}$ is the averaged total energy after the advection step (equation \ref{eq:Yadv}). The provisional temperature $T^{(adv)}$ is then obtained from $p^{(adv)}$ also by means of the EOS,
\begin{equation}
  T^{(adv)} = \frac{\overline{\left(1 - \rho b\right)}p^{(adv)} + \overline{\Pi(1 - \rho b)}}{\overline{\rho(c_p - c_v)}}.
	\label{eq::23}
\end{equation}
Equations \ref{eq::18} and \ref{eq::19} are then rearranged in the form of a Poisson-Helmholtz system of mutually coupled equations,
\begin{equation}
	\bnabla\bcdot\left(k\bnabla T^{n+1}\right) + \lambda_1 T^{n+1} + \lambda_2 p^{n+1} = \lambda_T,
	\label{eq::27}
\end{equation}
\begin{equation}
	\bnabla\bcdot\left(\frac{1}{\rho}\bnabla p^{n+1}\right) + \lambda_3 p^{n+1} + \lambda_4 \bnabla\bcdot\left(k\bnabla T^{n+1}\right) = \lambda_p.
	\label{eq::28}
\end{equation}
where,
\begin{align}
	\lambda_1 &= -\frac{\rho c_p}{\Delta t},\label{eq::29}\\
	\lambda_2 &= \frac{\beta T^{(adv)}}{\Delta t},\label{eq::30}\\
	\lambda_3 &= -\frac{1}{\Delta t^2}\left(\frac{\gamma}{\rho c^2} - \frac{\beta^2 T^{(adv)}}{\rho c_p}\right) ,\label{eq::31}\\
	\lambda_4 &= \frac{\beta}{\rho c_p},\label{eq::32}\\
	\lambda_T &= \lambda_1 T^{(adv)} + \lambda_2 p^{(adv)},\label{eq::33}\\
	\lambda_p &= \lambda_3 p^{(adv)} +\frac{1}{\Delta t}\bnabla\bcdot\boldsymbol{u}^*.\label{eq::34}
\end{align}
This system is of the form,
\begin{equation}
	\mathcal{L}(\boldsymbol{a}) = \boldsymbol{b},
\end{equation}
where, $\mathcal{L}(\bcdot)$  is a linear operator, and $\boldsymbol{a} = [T^{n+1} \quad p^{n+1}]^T$ and $\boldsymbol{b} = [\lambda_T \quad \lambda_p]^T$  are  both vectors. This system of mutually coupled equations can therefore be solved efficiently using a multigrid implicit solver, described for the SGN equations by Popinet \cite{jcp-popinet-2015}. When solving time-dependent problems, a good initial guess $\boldsymbol{\tilde{a}} = \boldsymbol{a} - \delta\boldsymbol{a}$, is available, where $\delta\boldsymbol{a}$ is an unknown correction. Therefore, it is usually more efficient to solve for the equivalent problem,
\begin{equation}
	\mathcal{L}(\delta\boldsymbol{a}) = \boldsymbol{b} - \mathcal{L}(\boldsymbol{\tilde{a}})  = \boldsymbol{res},
\end{equation}
where $\boldsymbol{res}$ is the residual. Owing to the linearity of the operator $\mathcal{L}(\bcdot)$, $\delta\boldsymbol{a}$ can be added to the initial guess $\boldsymbol{\tilde{a}}$, and the process is then repeated until the residual falls below a given tolerance. The procedure can be summarised by the following steps:
\begin{enumerate}
	\item Compute the residual $\boldsymbol{res} = \boldsymbol{b} - \mathcal{L}(\boldsymbol{\tilde{a}})$.
	\item If $\left\lVert\boldsymbol{res}\right\rVert < \epsilon$, $\boldsymbol{\tilde{a}}$ is good enough, stop.
	\item Else, solve $\mathcal{L}(\delta\boldsymbol{a})\simeq\boldsymbol{res}$.
	\item Add $\delta\boldsymbol{a}$ to $\boldsymbol{\tilde{a}}$ and go back to step 1. 
\end{enumerate}
For all the computations reported in this manuscript, the tolerance  $\epsilon$ is set to $10^{-6}$.  The multigrid solver therefore yields estimated values of $T^{n+1}$ and $p^{n+1}$ that can be readily use to compute the fluxes required to update the averaged momentum and the total energy. The velocity field $\boldsymbol{u}^{n+1}$, and by extension the momentum, is then computed using Eq. \ref{eq::unp1}.

To update the total energy of each component, we account for the fact that the resultant pressure at the cell centres corresponds to the one-fluid averaged field $p^{n+1} = C p_1^{n+1} + (1 - C)p_2^{n+1}$, out of which the pressure in each phase is derived using Laplace's law,
\begin{align}
	p_1^{n+1} &= p^{n+1} + (1 - C)\sigma\kappa^{n+1},\label{eq::37}\\
	p_2^{n+1} &= p^{n+1} - C\sigma\kappa^{n+1}.\label{eq::38}
\end{align}
These pressures are then used to update the total energy in each phase at the end of the time step,
\begin{equation}
	(C_i \rho_i e_{T,i})^{n+1} = (C_i \rho_i e_{T,i})^{(adv)} + \Delta t\left[
	- C_i \bnabla\bcdot\left( p_i \overline{\boldsymbol{u}}\right)^{n+1} 
	+ C_i \bnabla\bcdot\left( \mu_i\left(\bnabla \overline{\boldsymbol{u}} + \bnabla \overline{\boldsymbol{u}}^T\right) \bcdot \overline{\boldsymbol{u}}\right)^{n+1} 
	- C_i \bnabla\bcdot\boldsymbol{q}_i^{n+1}\right].
	\label{eq::39}
\end{equation}
Once all primitive variables are updated, it is possible to compute the final values of the derived variables such as the pressure and temperature fields via the EoS, which will the be consistent with the values of the conservative variables obtained at the end of the time-step. The numerical scheme is summarised in the commented algorithm \ref{alg}.

\RestyleAlgo{ruled}
\SetKwComment{Comment}{\textcolor{red}{/*} }{ \textcolor{red}{*/}}
\begin{algorithm}
	\caption{Summary of the algorithm (Comments in red)}\label{alg}
	Initialisation of $C_1\rho_1$, $C_2\rho_2$, $p$, $T$ (Eq. \ref{eq::23}), $\overline{\rho\boldsymbol{u}}$, $C_1\rho_1e_{T,1}$, $C_2\rho_2e_{T,2}$\Comment*[r]{\textcolor{red}{C\textsubscript{1}=C, C\textsubscript{2}=1-C}}
	\While{$t < t_{end}$}{
		Set $\Delta t$\Comment*[r]{\textcolor{red}{e.g. using an acoustic CFL}}
		Obtain $\boldsymbol{q}_i = C_i\rho_i\frac{\overline{\rho\boldsymbol{u}}}{C_1\rho_1 + C_2\rho_2}$\;
		Clone $\rho_i$, $q_i$ and $E_i$ as tracers to be advected with $C$\;
		Perform directional-split advection to evaluate $\boldsymbol{F}^{(adv)}$ \cite{jcp-fuster-2018}\;
		Compute $\overline{\rho\boldsymbol{u}}^{(adv)} = \boldsymbol{q}_1 + \boldsymbol{q}_2$\;
		Solve the mixture momentum equation for $\overline{\boldsymbol{u}}_{pred}^{n+1}$ (Eq. \ref{eq::2num})\;
		Define a provisional velocity field $\overline{\boldsymbol{u}}^*$ (Eq. \ref{eq::provel})\;
		Obtain face velocities\Comment*[r]{\textcolor{red}{Average of centre velocities of neighbouring cells}}
		Obtain provisional pressure $p^{(adv)}$ (Eq. \ref{eq::20})\Comment*[r]{\textcolor{red}{EOS}}
		Obtain provisional temperature $T^{(adv)}$ (Eq. \ref{eq::23})\Comment*[r]{\textcolor{red}{EOS}}
		Evaluate $\beta$ (Eq. \ref{eq::10})\;
		Evaluate $\rho c^2$ (Eq. \ref{eq::11})\;
		Evaluate $\lambda$ coefficients (Eqs. \ref{eq::29}--\ref{eq::33})\Comment*[r]{\textcolor{red}{Previously computed face velocities are used to evaluate the divergence term in Eq. \ref{eq::34}}}
		Solve the Poisson-Helmholtz system of mutually coupled equations \ref{eq::27}--\ref{eq::28} for $p^{n+1}$ and $T^{n+1}$\;
		Update face velocities (Eq. \ref{eq::17})\Comment*[r]{\textcolor{red}{Cell centre velocities, which are used to evaluate the momentum, are computed from face velocities by the same averaging method}}
		Update $C_1\rho_1e_{T,1}$ (Eq. \ref{eq::39}) using $p_1^{n+1}$ (Eq. \ref{eq::37})\;
		Update $C_2\rho_2e_{T,2}$ (Eq. \ref{eq::39}) using $p_2^{n+1}$ (Eq. \ref{eq::38})\;
	}
\end{algorithm}

\section{Test cases}\label{sec::4}
In this section, we propose test cases used to validate the correct implementation of the thermal effects in compressible solvers in both linear and strongly non-linear regimes. The results are compared either to classical numerical methods and models already available in the literature, and to analytical solutions when available.
\subsection{Epstein-Plesset like problem for temperature}\label{sec::4.1}
The first test case is inspired by Epstein and Plesset (1950) \cite{jcp-epstein-plesset-1950} who came up with analytical solutions for the shrinkage and growth of gas bubbles in undersaturated and supersaturated liquid-gas solutions, respectively. The change in bubble radius is driven by the diffusion of gas across the interface, given an initial difference between the gas concentration at the interface and in the liquid bulk. An analytical solution is reached provided one neglects the advective terms in the diffusion equation. This assumption is physically justified when the diffusive process is slow, which is typically the case for gas diffusion in liquids for small gas concentrations, thus avoiding convective effects by density gradients \cite{jfm-enriquez-2014}.

In this paper we test the shrinkage and growth of a spherical gas bubble in a liquid due to the diffusion of temperature between the two phases. From a quick look at the equation of state for an ideal gas, one can notice that the gas expands if heated, with its pressure kept the same, and shrinks if cooled. Even before any formal statement of an equation of state was put forth, this behaviour had been observed. Charles performed the first experiments in 1787, credited later on by Gay-Lussac in 1802 who published the linear relationship between volume and temperature at constant pressure \cite{gay-lussac-1802}. This was later called Charles's law,
\begin{equation}
	\frac{V_1}{T_1} = \frac{V_2}{T_2}.
	\label{charles}
\end{equation}
We perform axisymmetric simulations of an air bubble inside liquid water, subject to an initial temperature difference between the gas and the liquid. The parameters of this problem, as well as the properties of the fluids are rendered dimensionless by the liquid density $\rho_l = 975.91~kg/m^3$, pressure $p_\infty = 5\times 10^6 ~Pa$, temperature $T_\infty = 350~K$ and initial bubble radius $R_0 = 10^{-4}~m$. The domain size is set to $L = 8R_0$. A key point is to set a uniform pressure $p_\infty$ across the whole domain. For the sake of simplicity, we neglect surface tension so as to avoid dealing with the Laplace pressure jump, which induces pressure changes inside the bubble as it shrinks or expands. This will also simplify the theory to which we will compare our numerical results. However, viscosity is taken into account for it damps any interfacial corrugation that might arise from the absence of capillary forces.
\begin{figure}
	\centering\includegraphics[width=\textwidth]{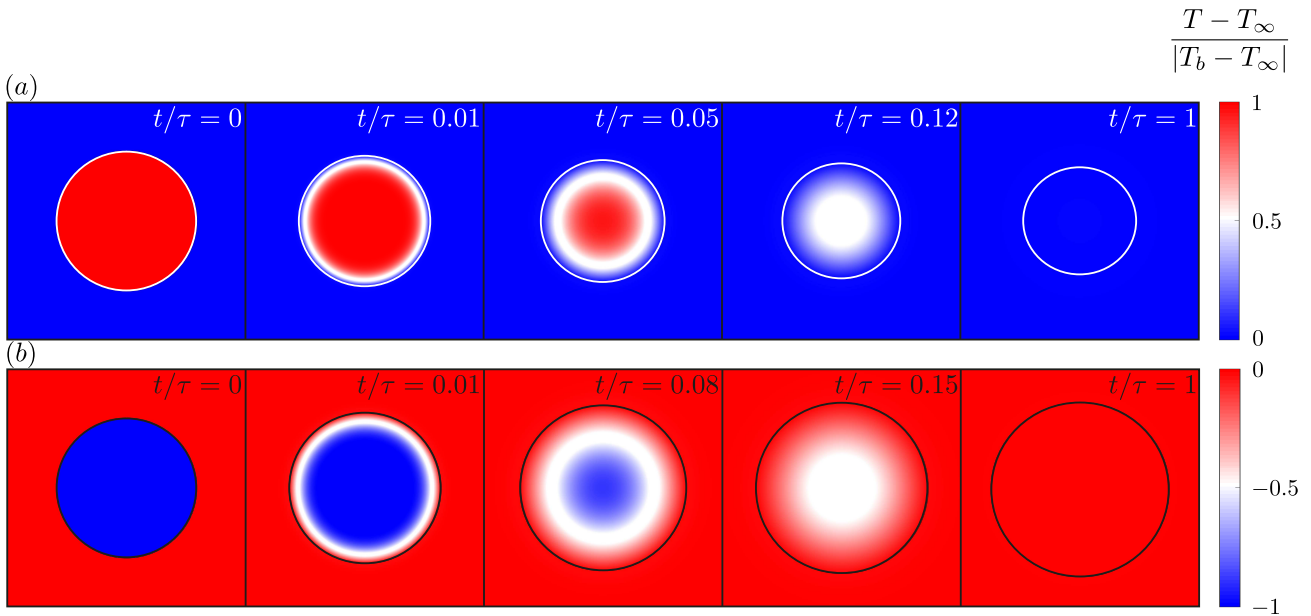}
	\caption{(a) A gas bubble, initially hotter than the surrounding liquid $\Delta T = T_b - T_\infty = 350~K$, shrinks and reaches a new radius while approaching the thermal equilibrium with the liquid. (b) A gas bubble, initially cooler than the surrounding liquid $\Delta T = T_b - T_\infty = -175~K$, expands and reaches a new radius while approaching the thermal equilibrium with the liquid. The colour code depicts the temperature field, and $\tau = R_0^2/\kappa_{th,g}$ is the diffusive time scale of the gas.}
	\label{fig::1}
\end{figure}
Figure \ref{fig::1}a presents a sequence of events for a case where the bubble is initially hotter than the liquid, with $\Delta T = T_b - T_\infty = 350~K$. The time is shown in multiples of the diffusive time scale of the gas $\tau = R_0^2/\kappa_{th,g} $, with $\kappa_{th,g} = k_g/\rho_g c_{p,g}$ the thermal diffusivity of the gas. One clearly sees that as the heat diffuses into the liquid, the bubble shrinks until it reaches a new equilibrium state. Owing to the much higher liquid thermal conductivity, one hardly sees any increase in the liquid temperature, even at the bubble wall. Heat is rapidly diffused and the assumption of a constant temperature at the bubble interface is usually a decent approximation in theoretical models. This is mainly the case for non-condensible gas bubbles in sufficiently cold liquids \cite{jasa-prosperetti-1988}. Figure \ref{fig::1}b shows the case where the bubble is initially cooler than the liquid, with $\Delta T = T_b - T_\infty = -175~K$. Heat thus diffuses into the bubble interior which keeps expanding until thermal equilibrium with the liquid is established.

\begin{figure}
	\centering\includegraphics[width=\textwidth]{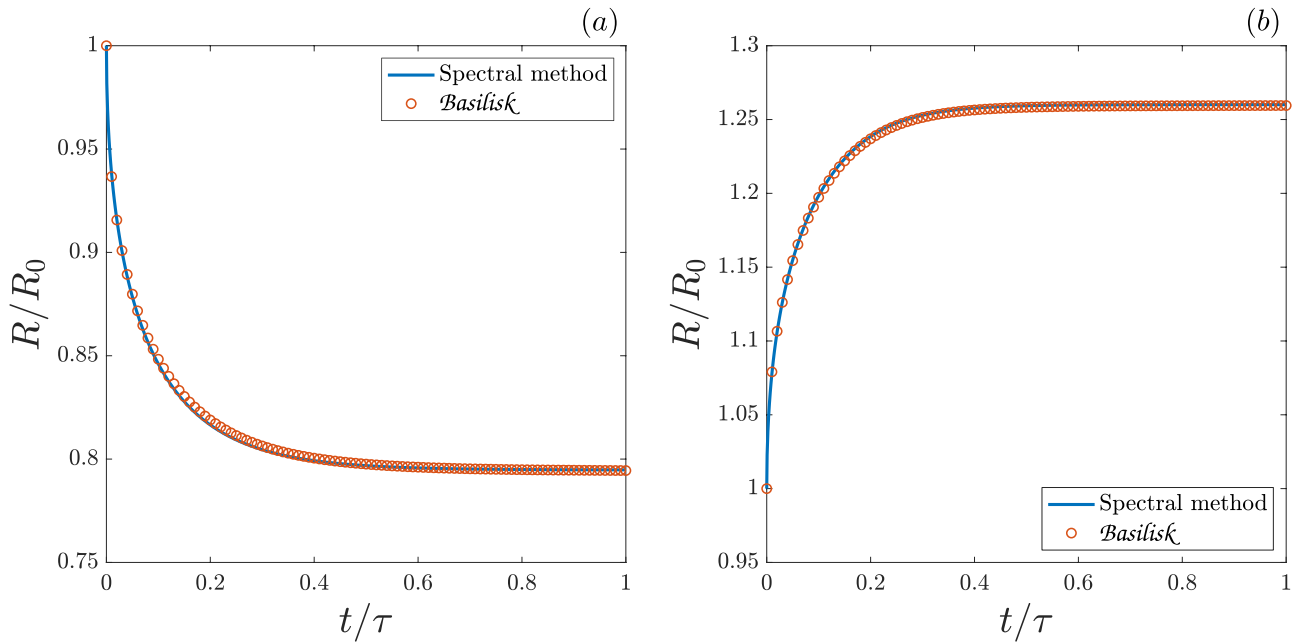}
	\caption{A comparison between our code, implemented in {\fontfamily{qzc}\selectfont Basilisk}, and the spectral method's solution for the cases(a) $\Delta T = T_b - T_\infty = 350~K$ and (b)  $\Delta T = T_b - T_\infty = -175~K$.}
	\label{fig::2}
\end{figure}

The equilibrium radii for both cases are given by equation \ref{charles}. The temporal evolution of the bubble radius is of interest as well and should be validated. As previously mentioned, Epstein and Plesset (1950) analytically derived $R(t)$ by neglecting the advective terms; however, making the same approximation in this case proved to be a simplistic solution. The reason is that this process is relatively fast, especially at early times where $\dot{R}$ is of large magnitude. Therefore, the advective terms could not be simply neglected. To make sure that our code also captures the correct temporal evolution of $R(t)$, we recur to a spectral method solution of the enthalpy equation inside the gas, coupled to an equation for the bubble radius \cite{pof-hao-1999a,pof-hao-1999b,jasa-stricker-2011,prf-hao-2017}. In the absence of viscous dissipation, the enthalpy equation inside the bubble, with temperature as the primitive variable, is written as \cite{jfm-prosperetti-1991}
\begin{equation}
	\frac{\gamma}{\gamma - 1}\frac{p}{T}\left[\frac{\p T}{\p t} + \frac{1}{\gamma p}\left(\left(\gamma - 1\right)k_g\frac{\p T}{\p r} - \frac{1}{3}r\dot{p}\right)\frac{\p T}{\p r}\right] - \dot{p} = k_g\nabla^2 T,
	\label{enthalpy}
\end{equation}
with
\begin{equation}
	\dot{p} = \frac{3}{R}\left[\left(\gamma - 1\right)k_g\frac{\p T}{\p r}\biggr\rvert_R - \gamma p \dot{R}\right],
	\label{radiusT}
\end{equation}
the equation  of pressure, assumed to be spatially uniform inside the bubble, which is typically the case for low densities of the gas \cite{ijmf-li-2021}. In our case, $\dot{p} = 0$ since the motion is driven by a temperature rather than by a pressure gradient. Equation \ref{radiusT} is thus simplified to a description of the temporal evolution of the bubble radius as a function of the temperature gradient at the interface. As can be seen, at $t = 0$, $\dot{R}\rightarrow\infty$. The assumption of a constant temperature at the bubble interface is employed, enabling us to solve the coupled system of equations  \ref{enthalpy}--\ref{radiusT} only, completely disregarding what happens in the liquid. The details of the spectral method \cite{jasa-stricker-2011} used for the solution is described in \ref{appA}.

Figure \ref{fig::2} shows a pretty good agreement between our current implementation of the thermal effects and the spectral method's solution. Our code well predicts the equilibrium radii as well as the temporal evolution in both cases. There is a slight discrepancy in $R(t)$, most probably due to the fact that viscosity is taken into account in our code. In the spectral method, an inviscid liquid was assumed. Viscosity dampens the motion and this is why we see a very small offset between the two solutions, with ours being slightly slower. Were viscosity to be included in the spectral solution, an additional equation, of the Rayleigh-Plesset type, would have been needed for $R(t)$. But this approximation is perfectly sufficient for our current purposes.

\begin{figure}
	\centering\includegraphics[width=\textwidth]{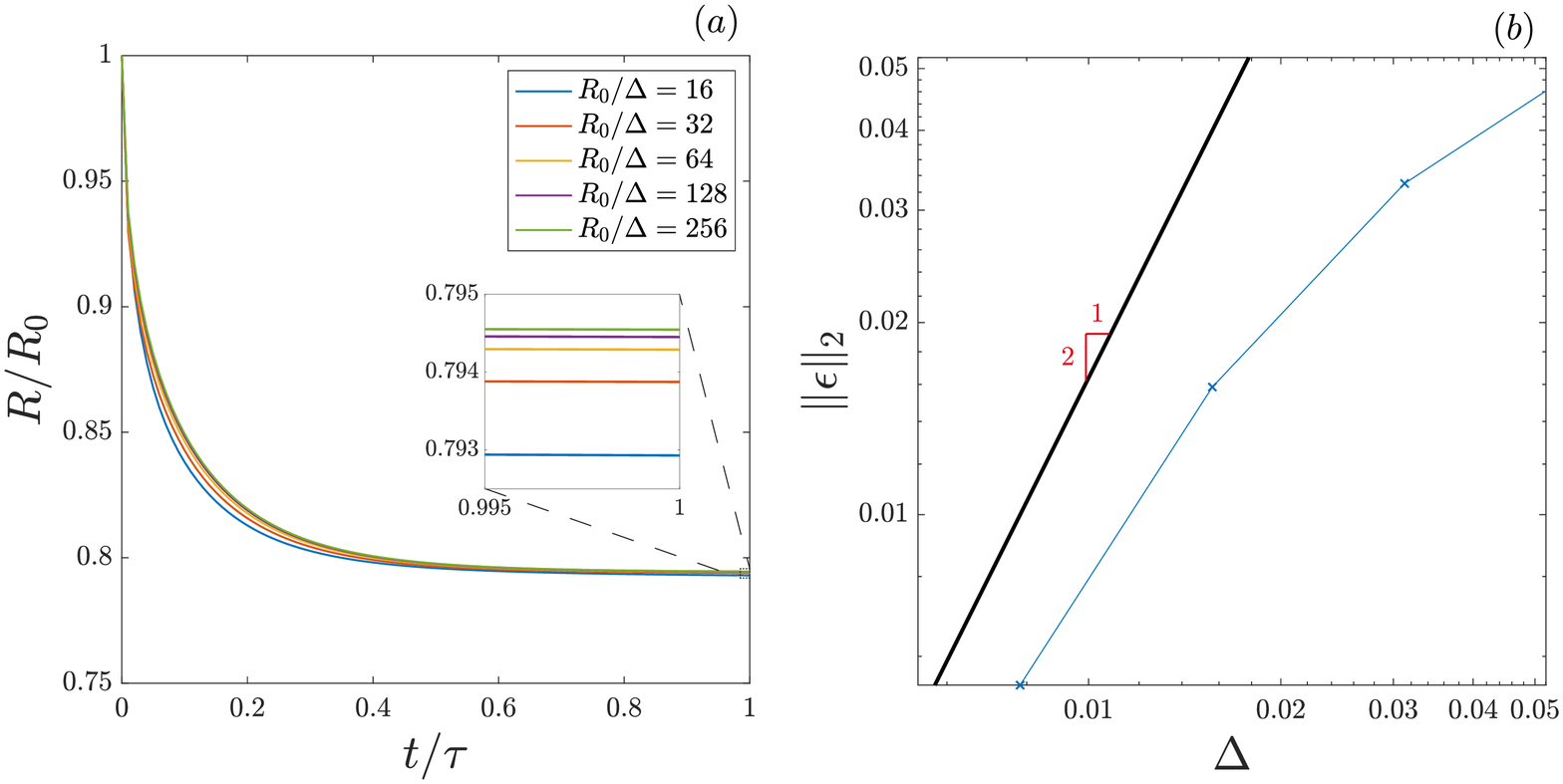}
	\caption{(a) A grid convergence study for the cast $\Delta T = T_b - T_\infty = 350~K$. The inset is a zoom on $R/R_0$  for $t\simeq\tau$.  The different curves are for different numbers of cells per initial bubble radius. (b) A log-log plot of the $L^2$ norm of the error as a function of the mesh size. The black line, with a slope equal to 2, depicts a numerical scheme that is exactly second order convergent in space.}
	\label{fig::3}
\end{figure}
Figure \ref{fig::3}a shows that as the grid is refined, $R(t)$ converges. 128 cells per initial bubble radius seems to be a sufficient resolution for decent results, since one barely discerns any changes with respect to a finer mesh. The results in figure \ref{fig::2} are thus produced with this resolution. Figure \ref{fig::3}b shows the $L^2$ norm of error with respect to the grid size, both in a log scale. Although our implementation of the thermal effects is done implicitly, meaning it being unconditionally stable, a constant ``diffusive" CFL is set for the grid convergence study,
\begin{equation}
	C = \kappa_{th,g}\frac{\Delta t}{\Delta^2} = \frac{1}{2}.
	\label{CFL}
\end{equation}
The $L^2$ norm of error is then computed from $R(t)$ as follows,
\begin{equation}
	\lVert\epsilon\rVert_2 = \left[\sum_{i = 0}^{100}
	\left(R_\Delta(i\tau/100) - R_{\Delta_*}(i\tau/100)\right)^2\right]^{1/2},
	\label{L2}
\end{equation}
where $\Delta_* = 1/256$ is the most refined mesh size. As expected, the smaller the $\Delta$, the smaller the error. Compared to the solid black line in figure \ref{fig::3}b, our method converges at second order in space.

\subsection{Free linear oscillations of a gas bubble}\label{sec::4.2}
Except in idealised analytical setups, every oscillating bubble experiences damping of its motion via several mechanisms: acoustic radiation in the liquid, viscous dissipation at the bubble wall, and thermal effects. These mechanisms alter the natural frequency of a bubble \cite{pm-minnaert-1933}. Chapman and Plesset (1971) \cite{chapman-plesset-1971} made available a theory for the free oscillations of a gas bubble in the linear regime. They quantify the contribution of each of the above mechanisms to the overall damping of the motion. In addition they show that as the equilibrium radius of the bubble decreases, the oscillation transitions from an adiabatic to an isothermal regime, in the framework of polytropic processes. Later, Prosperetti and co-workers studied the thermal effects in forced radial oscillations \cite{arfm-plesset-1977,jasa-prosperetti-1977,jasa-prosperetti-1988}. More recently, the theory was generalised, taking into account the effect of mass transfer on the attenuation of the bubble motion \cite{jfm-fuster-2015,ijhmt-bergamasco-2017}.  In the present work, we will focus on the thermal conduction contribution to the damping of free linear oscillations of a gas bubble, and check that our code well captures the predicted transition. For that end, both viscosity and surface tension are neglected.

We briefly describe the theory used for the comparison. For the details, the reader is referred to the relevant publication \cite{chapman-plesset-1971}.
The temporal evolution of the bubble radius is described as
\begin{equation}
	R(t) = R_0 + \xi\exp{\Omega t} = R_0 + \xi\exp{\left[\left(\zeta + i\omega\right)t\right]},
	\label{Reps}
\end{equation}
where $R_0$ is the equilibrium radius, $\xi\ll R_0$ is the amplitude of the perturbation, $\zeta$ a damping factor and $\omega$ the angular frequency of the oscillation.

By linearising the equations, i.e. conservation of mass, momentum and energy, as well as the equation of state, and by applying the linearised boundary conditions, Chapman and Plesset (1971) \cite{chapman-plesset-1971} were able to find an equation for $\Omega$ which, in the absence of capillary and viscous effects, reads
\begin{equation}
	\Omega^2 = -\left(1 + \frac{\Omega R_0}{c_l}\right)\frac{G}{\rho_lR_0^2},
	\label{eq::heq}
\end{equation}
where
\begin{equation}
	G = \frac{p_0\kappa_{th,g} R_0^2\left(\lambda_2 - \lambda_1\right)}{\left(\frac{\Omega}{\lambda_1} - \kappa_{th,g}\right)\left(R_0\lambda_1^{1/2}\coth R_0\lambda_1^{1/2}\right) - \left(\frac{\Omega}{\lambda_2} - \kappa_{th,g}\right)\left(R_0\lambda_2^{1/2}\coth R_0\lambda_2^{1/2}\right)},
\end{equation}
and where the constants $\lambda_1$ and $\lambda_2$ are the roots of the quadratic equation
\begin{equation}
	\Omega^2 - \left(\gamma\frac{p_0}{\rho_0} + \Omega\kappa_{th,g}\right)\lambda + \frac{p_0}{\rho_0}\frac{\kappa_{th,g}}{\Omega}\lambda^2 = 0.
\end{equation}
Equation \ref{eq::heq} can then be solved in the complex plane by any root finding algorithm. Once $\Omega$ is computed, the logarithmic decrement $\Lambda$, an indicator of the motion's attenuation, can be readily obtained as
\begin{equation}
	\Lambda = \frac{\zeta}{f} = 2\pi\frac{\zeta}{\omega}.
	\label{logdec}
\end{equation}
where $f$ is the frequency of the oscillation. Minnaert (1933) \cite{pm-minnaert-1933} derived the natural frequency of a bubble in an adiabatic regime. In such cases, the polytropic coefficient is equal to the ratio of specific heats in the gas $\gamma_p= \gamma$, and the natural frequency, again in the absence of viscous, capillary and acoustic effects, reads
\begin{equation}
	\omega_0 = \left(\frac{3\gamma p_0}{\rho_l R_0^2}\right)^{1/2},
	\label{minnaert}
\end{equation}
where $p_0$ is the equilibrium pressure. Since heat transfer between the phases is taken into account, equation \ref{minnaert} no longer holds. Instead, an ``effective" polytropic coefficient $\gamma_p$ is computed as a correction to Minnaert's frequency, in order to include thermal effects,
\begin{equation}
	\omega = \left(\frac{3\gamma_p p_0}{\rho_l R_0^2}\right)^{1/2}.
	\label{kappaeff}
\end{equation}

\begin{figure}
	\centering\includegraphics[width=\textwidth]{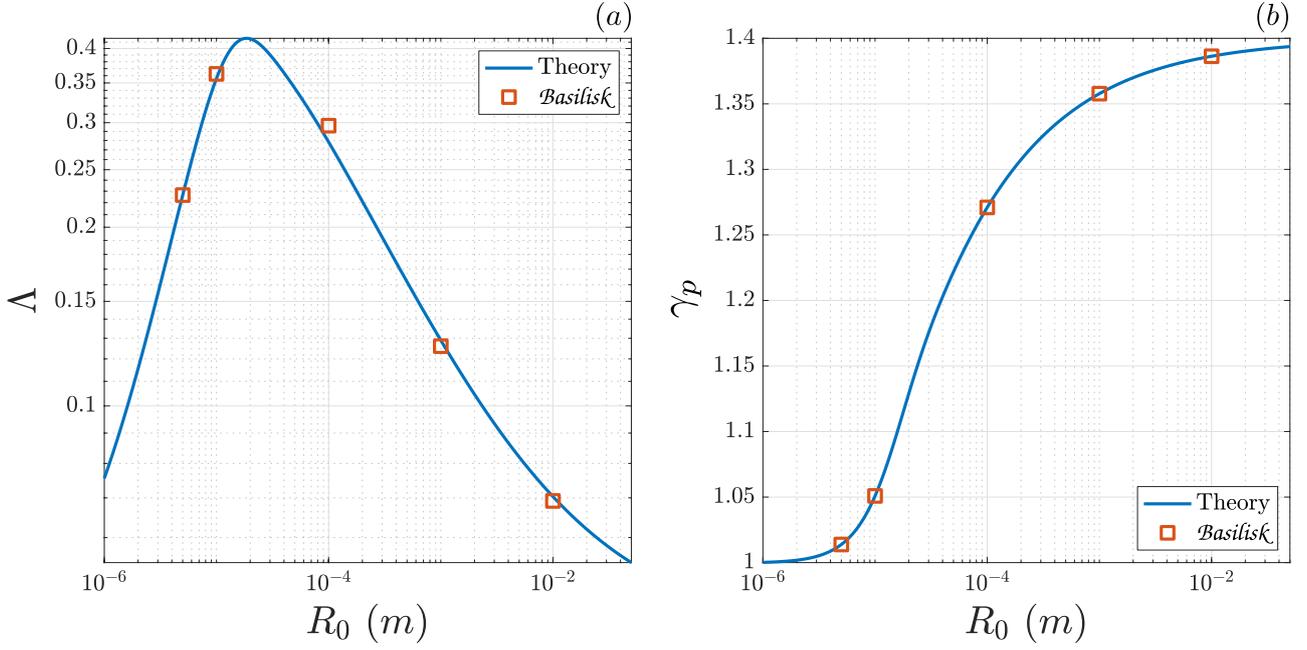}
	\caption{(a) Logarithmic decrement at the end of one cycle of free bubble oscillation for different values of the equilibrium radius. (b) Effective polytropic coefficient $\gamma_p$ computed using equation \ref{kappaeff} for different values of the equilibrium radius.}
	\label{fig::4}
\end{figure}
\begin{figure}
	\centering\includegraphics[width=\textwidth]{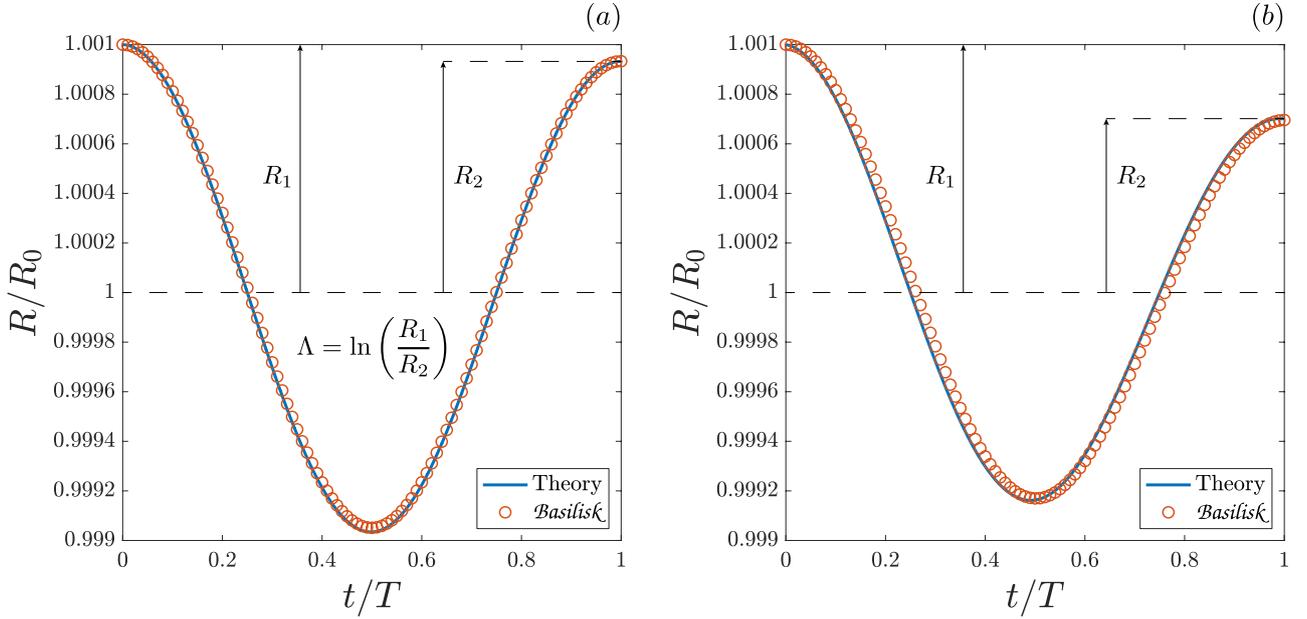}
	\caption{(a) Normalised radius as a function of time for  (a) $R_0 = 10^{-2} \ m$ and (b) $R_0 = 10^{-5} \ m$. Times are normalised with the respective oscillation period for each case $T = 1/f = 2\pi/\omega$. The solid lines are obtained using equation \ref{Reps}.}
	\label{fig::5}
\end{figure}

We perform spherically symmetric simulations of an air bubble in liquid water at $T_\infty = 350 \ K$ and $p_\infty = 1 \ atm$. The equilibrium density of air is computed using the ideal gas equation of state $\rho_0 = p_0/\left[T_0\left(c_{p,g} - c_{v,g}\right)\right]$, where $p_0 = p_\infty$ in the absence of surface tension and $T_0 = T_\infty$. The bubble is slightly put out of equilibrium with the initial radius $R_i = 1.001R_0$. Since mass is conserved, the initial density is given by $\rho_i = \rho_0\left(R_0/R_i\right)^3$. The initial pressure is computed assuming a polytropic process, with its coefficient given by equation \ref{kappaeff},
\begin{equation}
	p_i = p_0\left(\frac{R_0}{R_i}\right)^{3\gamma_p}.
\end{equation}
The initial temperature $T_i$ is then computed using the equation of state. The thermodynamic properties of both fluids are constants taken at the equilibrium pressure and temperature. The domain size is set to $L = 256R_0$, large enough for a bubble to complete an oscillation cycle before being affected by spurious pressure reflections at the boundary from previous acoustic emissions, $2L > 2\pi c_l/\omega$. The resolution is set to 2048 cells per equilibrium bubble radius ($\Delta = R_0/2^{11}$). For a correct prediction of the acoustic emissions, the pressure waves should be accurately resolved. Therefore, an acoustic CFL condition is employed, based on the speed of sound in the liquid,
\begin{equation}
	C_{ac} = c_l\frac{\Delta t}{\Delta} = \frac{1}{2}.
\end{equation}

Simulations were carried out for $R_0 \in \{5\times10^{-6},10^{-5},10^{-4},10^{-3},10^{-2}\} \ m$. Figure \ref{fig::4}a shows the theoretical logarithmic decrement (equation \ref{logdec}). {\fontfamily{qzc}\selectfont Basilisk }'s results computed at the end of one free oscillation cycle are in good agreement with the theory. This means that the code correctly captures the thermal damping.  Figure \ref{fig::4}b shows a perfect agreement between the theoretically and numerically computed effective polytropic coefficients. The code correctly captures the transition from an adiabatic to an isothermal oscillation as the equilibrium radius decreases. Air as a diatomic gas has $\gamma = 1.4$, so the effective $\gamma_p$ decreases from 1.4 to 1.

Figure \ref{fig::5}a shows the temporal evolution of the radius for the case $R_0 = 10^{-2} \ m$, compared with equation \ref{Reps}. This is an adiabatic regime as can be seen from figure \ref{fig::4}b. The damping merely consists of acoustic radiation. Very good agreement is achieved with the analytical solution, both in terms of the attenuation and the oscillation frequency. The figure also shows how the logarithmic decrement $\Lambda$ is extracted from the numerical simulations. Figure \ref{fig::5}b shows the comparison for the case $R_0 = 10^{-5} \ m$. This is a nearly isothermal case, and one can see that the motion is damped further. Thermal conduction now has an important contribution as compared to figure \ref{fig::5}a. The agreement is also good.

\subsection{Single bubble sonoluminescence (SBSL)}
Single bubble sonoluminescence is the periodic light emission from an acoustically strongly driven gas bubble at a specific set of parameters, i.e. forcing amplitude, frequency, concentration of dissolved gas etc \cite{rmp-brenner-2002}. The bubble strongly and rapidly collapses so that the internal energy is highly focused in a very small volume, leading to strong heating of the gas, partial ionisation, and a recombination of ions and electrons (thermal bremsstrahlung \cite{nature-hilgenfeldt-1999}). This process of light emission is surprisingly stable and periodic, and also visible to the naked eye in the dark \cite{rmp-brenner-2002}. It is a challenging problem from the numerical point of view for it is strongly non-linear, and an interplay between many physical aspects, i.e. heat and mass transfer, acoustic radiation etc. Since our numerical method does not allow mass transfer at the moment, this aspect will be neglected, despite being important experimentally and having a direct effect on the temperatures generated inside the bubble. We will assume that the bubble exists at the centre of a spherical flask of radius $R_\infty = 0.01 \ m$, surrounded by liquid water at atmospheric pressure.
\begin{figure}
	\centering\includegraphics[width=0.6\textwidth]{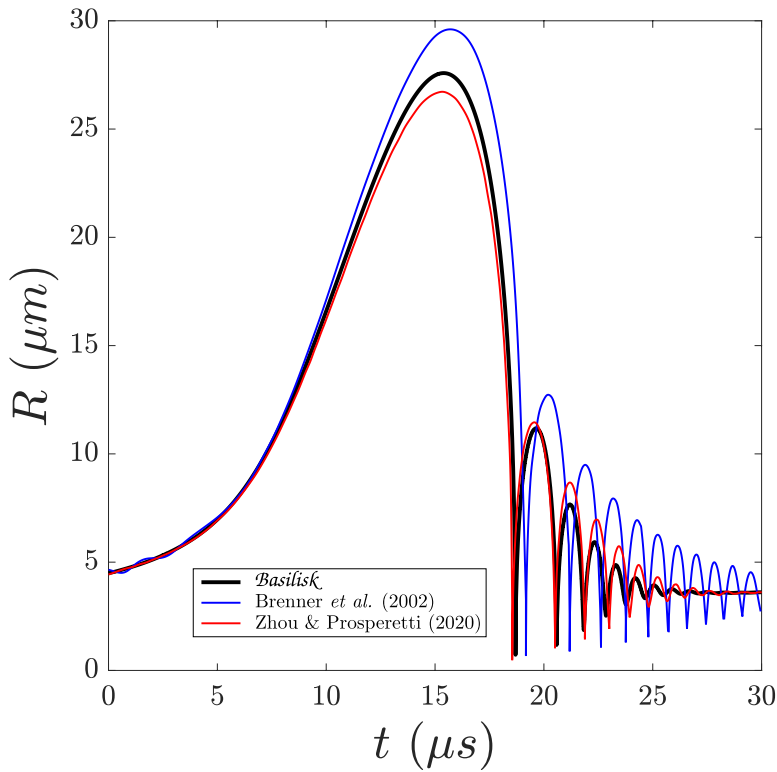}
	\caption{A single bubble of initial radius $R_0 = 4.5 \ \mu m$ at $p_\infty = 1 \ atm$ is subject to a pressure signal of amplitude $P_a = 1.2 \ atm$ at $r = 0$ and of frequency $f = 26.5 \ kHz$. The figure shows the bubble radius with respect to time, for different numerical methods.}
	\label{fig::7}
\end{figure}

After having verified that our code is capable of producing a standing wave in pure liquid \cite{jasa-fuster-2011} (\ref{appB}), we performed the simulation of a single sonoluminescent bubble in liquid water. The test case is inspired by Brenner, Hilgenfeldt \& Lohse (2002) \cite{rmp-brenner-2002} and is that of an Argon bubble  with an initial radius of $R_0 = 4.5 \ \mu m$, in a spherical flask of radius $R_\infty = 0.01 \ m$, driven with a pressure signal of amplitude $P_a = 1.2 \ atm$ and of frequency $f = 26.5 \ kHz$. Spherical symmetry is assumed, and the simulation is performed in the $r$-coordinate only. To achieve a standing wave of amplitude $P_a = 1.2 \ atm$ at $r = 0$, the amplitude $\Delta p_\infty$ of the sinusoidal driving is computed using equation \ref{swave}, and plugged in the Dirichlet boundary condition \ref{dirichlet}. Argon is a monoatomic noble gas, so its ratio of specific heats is $\gamma = 5/3$. Both viscous and capillary effects are taken into account in this simulation.
Figure \ref{fig::7} shows the bubble radius as a function of time for one oscillation cycle. Our results are compared to those of Brenner, Hilgenfeldt \& Lohse (2002) \cite{rmp-brenner-2002} and Zhou \& Prosperetti (2020) \cite{jfm-zhou-2020} for the same case. The former authors employed a Rayleigh type equation to obtain their result (blue dots in figure \ref{fig::7}), while the latter authors performed DNS of the bubble interior, including an equation for temperature, coupled with a Keller-Miksis equation for a description of the bubble radius (red dots in figure \ref{fig::7}). The bubble first expands isothermally, then violently collapses. Light is emitted at the end of this rapid adiabatic collapse. Afterbounces also occur until a new oscillation cycle begins. The results show reasonable agreement, particularly between the current and Zhou \& Prosperetti's (2020) \cite{jfm-zhou-2020} because both take into account thermal dissipation. Therefore, the results of both show further damping (especially for the afterbounces) than Brenner, Hilgenfeldt \& Lohse's (2002) \cite{rmp-brenner-2002} who treated the thermal damping only approximately. The current method is a full DNS, both inside and outside the bubble. This is why we also see further damping in the present results than in Zhou \& Prosperetti's (2020) \cite{jfm-zhou-2020}. The model they use considers the liquid to be only weakly compressible, so the amount of the bubble's internal energy lost to acoustic radiation is underestimated \cite{ijmf-li-2021}.

\begin{figure}
	\centering\includegraphics[width=\textwidth]{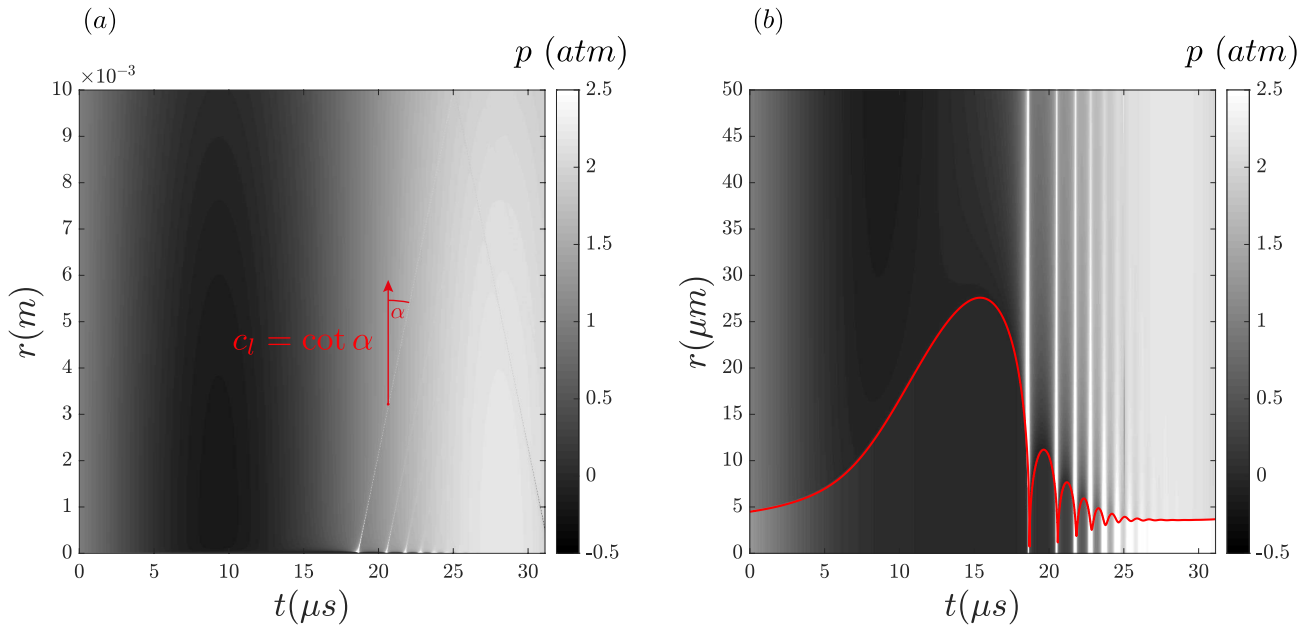}
	\caption{(a) Pressure field $p(r,t)$ inside the spherical flask with respect to time. (b) Zoom in on the radial range $[0-50] \ \mu m$, occupied by the oscillating bubble, the radius of which is denoted by the red line.}
	\label{fig::8}
\end{figure}

Figure \ref{fig::8}a shows the pressure field $p(r,t)$ inside the spherical flask with respect to time. On the large scale, one clearly recognizes the sinusoidal driving in the colour change from dark to light. At different times, one sees the propagation of high pressure waves as white straight lines, the slope of which is the speed of sound in the liquid, as indicated in the figure. A disadvantage of the present code in simulating acoustically driven bubbles is the reflection of emitted pressure/shock waves at the boundary which spuriously contaminate the physical process. This is why only one oscillation cycle is simulated, with a numerical domain larger than 2000 times the initial radius of the bubble. With a careful inspection, one sees that around $t = 25 \ \mu s$, the first emitted shock wave is reflected at $r = R_\infty$, and is carried back as a rarefaction wave. However, since the numerical domain is large enough, the physical process remains intact, which would not be the case had the simulation been continued for an additional oscillation cycle. Figure \ref{fig::8}b is a zoom in on the region of interest, occupied by the oscillating bubble. The bubble radius is depicted by the red line, and one clearly sees that the pressure/shock waves that we just discussed are emitted at the moment of the main collapse, as well as at subsequent collapses from the afterbounces. The capturing of this acoustic dissipation is one of the advantages of this method over others \cite{rmp-brenner-2002,jfm-zhou-2020}.

\begin{figure}
	\centering\includegraphics[width=0.6\textwidth]{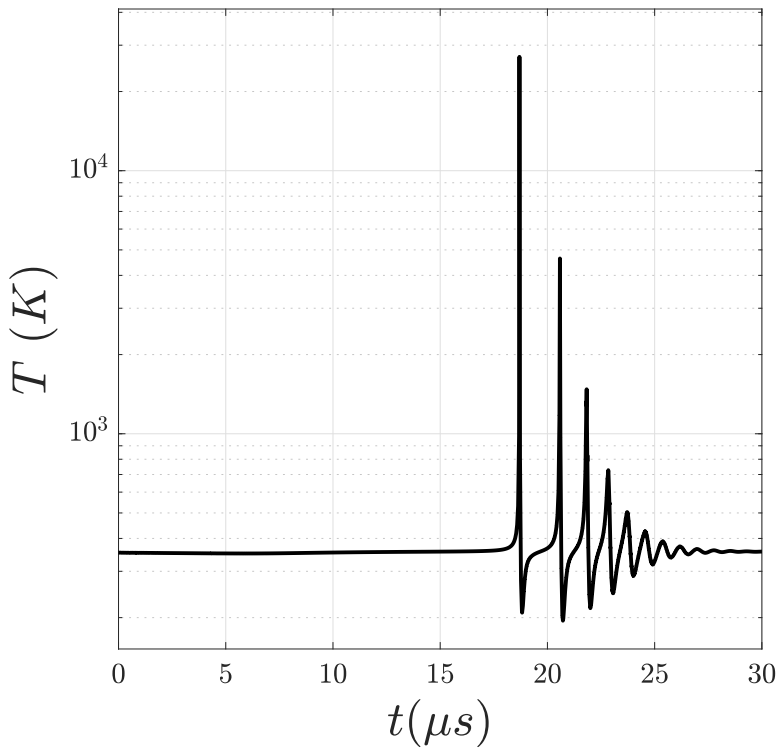}
	\caption{Bubble centre temperature $T(r = 0,t)$ as a function of time. The temperature is represented in a log scale.}
	\label{fig::9}
\end{figure}
Figure \ref{fig::9} shows the temperature, in log scale, at the bubble centre $r = 0$ as function of  time. At the moment of the main bubble collapse, the temperature peaks and reaches $T \sim 27500 \ K$. This is somewhat higher than the $15000~K
$ often reported \cite{nature-lohse-2005}, but consistent order of magnitude wise. The reason for the lower temperature  in experiment lies in the water molecules  which enter the bubble and  reduce the effective polytropic exponents, and partial ionisation, which does the same \cite{prl-toegel-2000}. $T(r=0)$ also increases at each of the subsequent collapses, with the peaks gradually decreasing in value since the rebounds and collapses become weaker and weaker over time due to viscous, acoustic and thermal damping.

\section{Numerical examples} \label{sec::5}
\subsection{Spherical Rayleigh collapse}
In this section, we perform spherically symmetric simulations of the Rayleigh collapse problem \cite{pm-rayleigh-1917}. But rather, the content of our bubble is gaseous instead of being void. Due to an initially lower pressure $p_0$ inside the bubble, the latter collapses and performs many oscillation cycles before reaching an equilibrium state, provided damping mechanisms exist of course. Otherwise, the bubble keeps oscillating indefinitely. In the current framework, the aim is to check the effect of heat transfer on the collapse, as compared to a purely adiabatic case. In particular, we compare the maximum pressure reached by the bubble at the end of the first collapse phase. Concurrently, the bubble reaches its minimum volume which we also compare.

We start by theoretically predicting the behaviour of the bubble in the two limiting cases: the adiabatic and isothermal bubble response of a bubble in an inviscid incompressible liquid in the absence of surface tension. For this end, we write the conservation of the mechanical energy in the liquid,
\begin{equation}
	\frac{\p}{\p t}\left[\rho_l \frac{1}{2}\boldsymbol{u}_l^2\right] + \bnabla\bcdot\left[\rho_l\frac{1}{2}\boldsymbol{u}_l^2 \boldsymbol{u}_l\right] = - \bnabla\bcdot\left( p_l \bcdot\boldsymbol{u}_l\right),
\end{equation}
and we integrate over the whole liquid volume using Gauss's theorem,
\begin{equation}
	\frac{dE_k}{dt} = - \int \bnabla\bcdot\left( p_l \bcdot\boldsymbol{u}_l\right) dV_l = -\int p_l (\boldsymbol{u}_l \bcdot \boldsymbol{n}) dS,
	\label{Ekevol}
\end{equation}
where $E_k = \int \frac{1}{2} \rho_l \boldsymbol{u}_l^2 dV_l$. The liquid volume is enclosed between two surfaces where the pressure is uniform: the interface, where in absence of surface and viscous effects the pressure is equal to that of the bubble $p_b$, and the far away boundary where the pressure is assumed to be constant and equal to $p_\infty$. Thus, eq. \ref{Ekevol} can be readily integrated in time between $t=0$ and any arbitrary instant,
\begin{equation}
	E_k(t) = \int_{V_0}^{V_b(t)} (p_\infty - p_b) dV_b,
\end{equation}
where we have imposed that the initial kinetic energy in the liquid is zero. We have also used the mass conservation relation $\int (\boldsymbol{u}_l \bcdot \boldsymbol{n}) dS = \frac{d V_b}{dt}$. If we evaluate this expression for the particular time at which the bubble reaches its minimum volume and the liquid velocity becomes zero, then,
\begin{equation}
	p_\infty\left(V_{min} - V_0\right) - \int_{V_0}^{V_{min}}p_b dV_b = 0,
	\label{eq::balance2}
\end{equation}
which can be integrated for a known relation between the bubble pressure and volume. For instance, if we assume a polytropic processes then $p_bV_b^\gamma = p_0V_0^\gamma$ (adiabatic case), then
\begin{equation}
	p_\infty\left(V_{min} - V_0\right) + \frac{p_0V_0^\gamma}{\gamma - 1}\left(V_{min}^{1 - \gamma} - V_0^{1 - \gamma}\right) = 0,
	\label{eq::adiabatic}
\end{equation}
which is a non-linear equation where $V_{min}$ can thus be computed using any root finding algorithm. The maximum pressure $p_{max}$ achieved by the bubble at this instant is immediately found using the relation between pressure and volume imposed by a polytropic process. Analogously, in the isothermal limit $p_bV_b = p_0V_0$ and the resulting equation is
\begin{equation}
	p_\infty\left(V_{min} - V_0\right) - p_0V_0\ln\left(\frac{V_{min}}{V_0}\right) = 0.
	\label{eq::isothermal}
\end{equation}

\begin{figure}
	\centering\includegraphics[width=\textwidth]{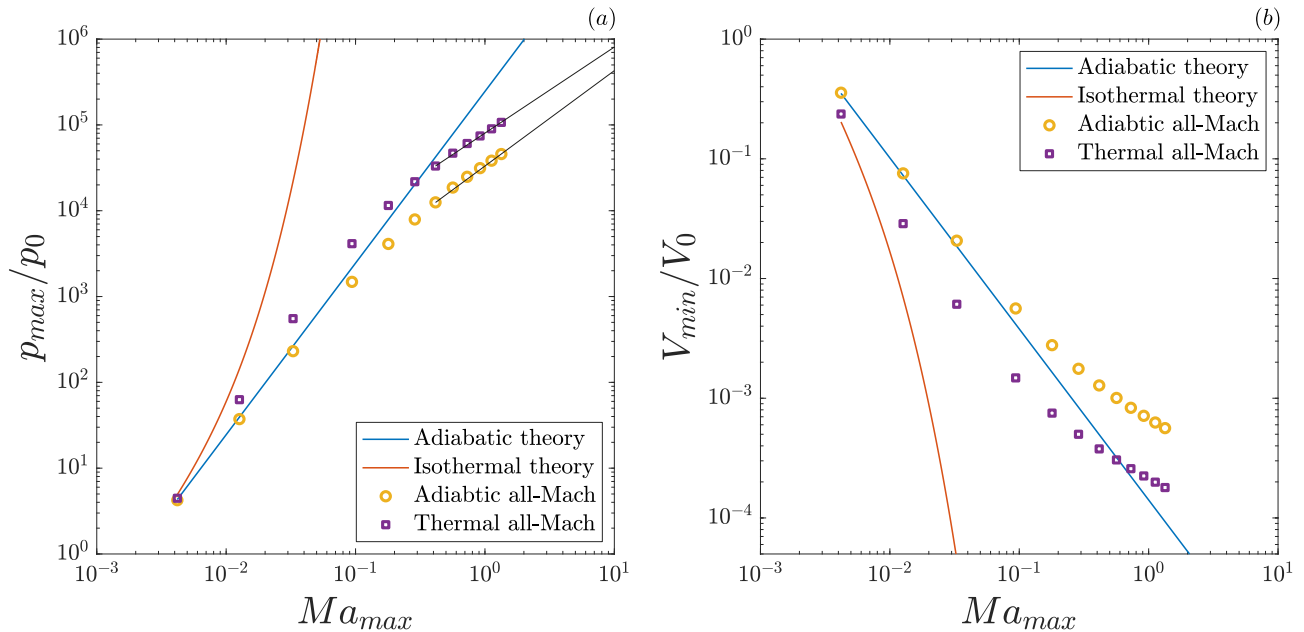}
	\caption{(a) Normalised maximum pressure achieved by the bubble at the end of its first collapse as a function of the maximum Mach number in the simulation. The adiabatic and isothermal theories are given by equations \ref{eq::adiabatic} and \ref{eq::isothermal}, respectively. (b) Normalised minimum volume reached by the bubble as a function of the maximum Mach number in the simulation.}
	\label{fig::new}
\end{figure}
For the simulations, an air bubble, with an initial radius $R_0 = 10^{-4}\ m$, is initialised with a pressure $p_0 = 0.1\ bar$. The surrounding liquid is assumed to be water at $T_\infty = 293.15\ K$, with a density $\rho_l = 998.21\ kg/m^3$. Simulations are done for a wide range of liquid far field pressures $p_\infty/p_0\in[2,100]$. In all cases, the bubble is initially supposed to be in thermal equilibrium with the surroundings $T_0 = T_\infty$. The domain size is set to 64 times the initial radius of the bubble, and the employed resolution is $\Delta = R_0/1024$.

Figure \ref{fig::new}a shows the normalised maximum pressure reached inside the bubble at the end of the first collapse phase, as a function of the maximum Mach number $Ma_{max} = U_{max}/c_l$, where $p_{max}^{ad} = p_0\left(V_0/V_{min}\right)^\gamma$ is the maximum pressure computed after obtaining $V_{min}$ using the adiabatic theory eq. \ref{eq::adiabatic}, and $U_{max} = \left(p_{max}^{ad}/\rho_l\right)^{1/2}$ given by the incompressible theory. As the initial pressure ratio $p_\infty/p_0$ increases, the collapse of the bubble becomes more violent and compressibility effects will become much more important. Therefore, a more descriptive dimensionless number is the aforementioned maximum Mach number. The result of the adiabatic simulations, where the thermal conductivities of both fluids are set to zero, are in agreement with the adiabatic theory for small $Ma_{max}$. As the latter number increases, the maximum pressure achieved in the simulations becomes less than that predicted by the incompressible adiabatic theory eq. \ref{eq::adiabatic}. A more important fraction of the bubble's internal energy is then radiated via pressure/shock waves to the surrounding liquid. Therefore, the bubble achieves less compression, and reaches minimum volumes that  are larger than those predicted by the theory, as can be seen in figure \ref{fig::new}b. When heat diffusion between the phases is taken into account, the bubble is compressed further. As expected, the results of the thermal all-Mach code in figures \ref{fig::new}a-b lie in the envelope delimited by both theories, until compressibility effects become prominent. At much larger pressure ratios, the Rayleigh collapses become much more violent, with a much higher collapse velocity $(p_\infty/\rho_l)^{1/2}$. Therefore the Peclet number, defined as $Pe = R_0(p_\infty/\rho_l)^{1/2}/\kappa_{th,g}$ where $\kappa_{th,g}$ is the thermal diffusivity of the gas, becomes much larger. This means that thermal diffusion happens at a much longer time as compared to advection. Therefore, thermal all-Mach should converge towards its adiabatic counterpart, which seems to be the trend in our simulations as well (solid black lines in figure \ref{fig::new}a). It must be stated that although the peak pressures reached during an isothermal compression are higher than those of an adiabatic one (figure \ref{fig::new}a), the internal energy $e = pV/\left(\gamma - 1\right)$ is smaller since the bubble volume reached at the end of the compression is smaller in the isothermal case (figure \ref{fig::new}b). Indeed, one would expect the internal energy to be smaller in the isothermal limit since heat is evacuated from the bubble.

\begin{figure}
	\centering\includegraphics[width=\textwidth]{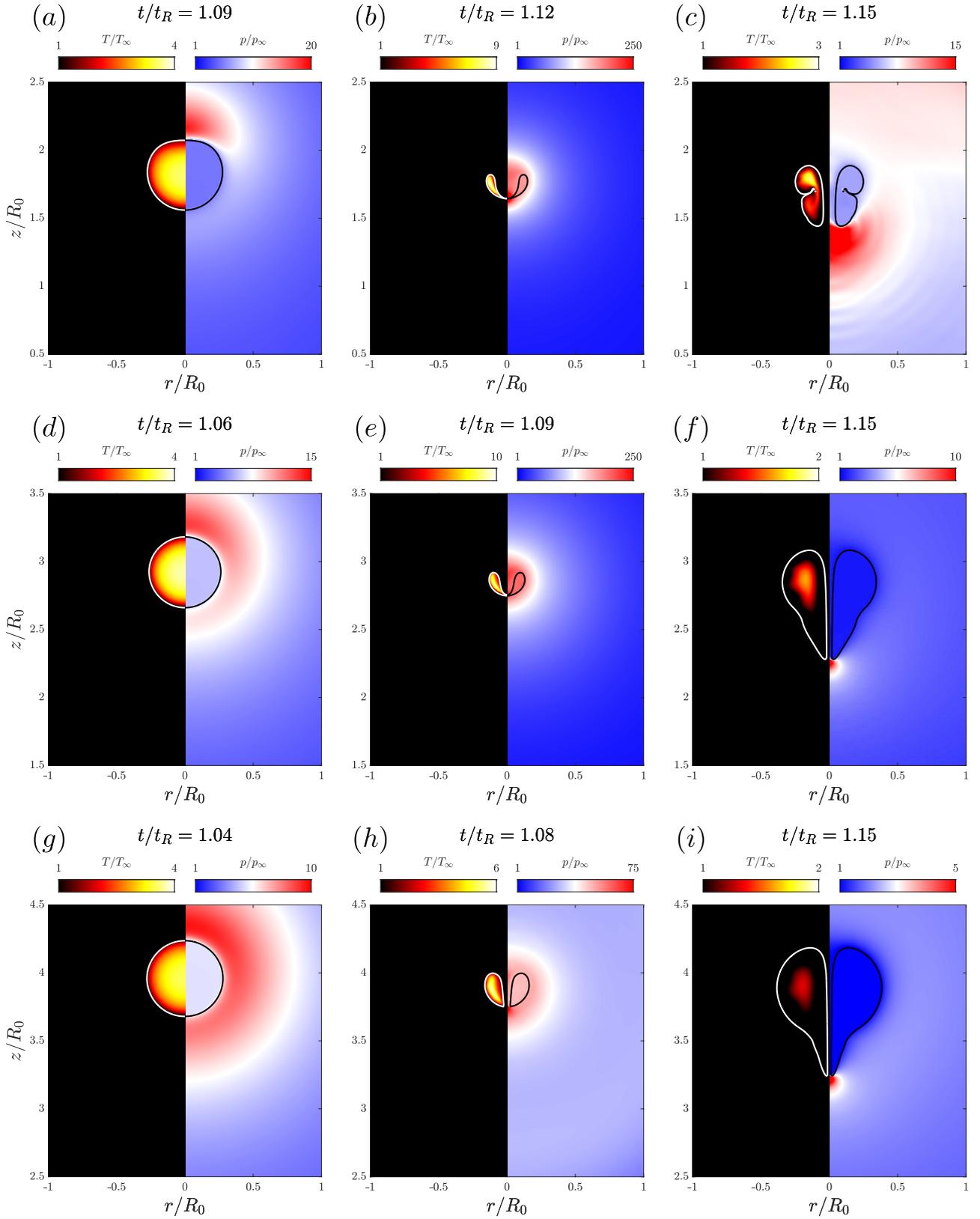}
	\caption{Snapshots of a collapsing bubble near a rigid boundary for (a-c) $\delta = 2$, (d-f) $\delta = 3$, (g-i) $\delta = 4$. The left hand panel of each snapshot shows the temperature field, whereas its right hand counterpart shows the pressure field. The respective instants in time are indicated on top of each snapshot.}
	\label{fig::10}
\end{figure}

\subsection{Bubble collapse near a rigid boundary}
Studies of a collapsing bubble near a rigid boundary abounds in the literature \cite{jcp-fuster-2018,jfm-popinet-2002,jfm-zeng-2018,jfm-beig-2018}. In the present work, we also tackle it, with a special focus on the temperature field and the heat flux across the bubble. Axisymmetric simulations of a collapsing air bubble in the vicinity of a rigid boundary are performed. The surrounding liquid is assumed to be water at $p_\infty = 1 \ atm$ and $T_\infty = 293.15 \ K$, with a density $\rho_l = 998.21 \ kg/m^3$. The bubble's radius is initially set to $R_0 = 10^{-4} \ m$, and its pressure to $p_{b,0} = p_\infty/20$. The parameters and properties of the problem are then rendered dimensionless by $R_0$, $\rho_l$, $\Delta p_0 = p_\infty - p_{b,0}$ and  $T_\infty$. Times will therefore be represented as multiples of the Rayleigh collapse time $t_R = 0.915R_0\left(\rho_l/\Delta p_0\right)^{1/2}$ \cite{pm-rayleigh-1917}. Water is supposed to be inviscid, with a hypothetical surface tension defined by a Weber number of $We = \Delta p_0R_0/\sigma = 1000$. The bubble is initially supposed to be in thermal equilibrium with the surroundings $T_{b,0} = T_\infty$, and with a density of $\rho_g = \rho_l/1000$. The domain size is set to 64 times the initial radius of the bubble, and the employed resolution is $\Delta = R_0/256$, gradually coarsened far from the bubble. Three cases are simulated, where the difference only lies in the initial distance $H$ between the bubble centre and the rigid wall ($z = 0$). Let $\delta = H/R_0$ be the stand-off ratio, it spans the following set $\delta\in\{2,3,4\}$. All other parameters are kept the same.

The collapse of the bubble is driven by the initial pressure difference $\Delta p_0$. As the bubble is compressed, its internal pressure increases. Lord Rayleigh (1917) \cite{pm-rayleigh-1917} was the first to quantify the huge increase of the liquid pressure at the bubble wall. The existence of a boundary breaks the spherical symmetry of the pressure field, which can be seen in figures \ref{fig::10}a, \ref{fig::10}d and \ref{fig::10}g where an imbalance in pressure exists between the top and bottom walls of the bubble. The closer the bubble is to the boundary, the more important this imbalance is. The effect of this can also be seen in the shape of the inner jet that pierces the bubble, directed towards the rigid boundary, and which becomes thicker for smaller $\delta$. The bubble reaches its minimum volume, associated with the highest internal pressure and temperature. The inner jet finally impacts the bottom wall of  the bubble and breaks it into a toroidal structure (figures \ref{fig::10}c, \ref{fig::10}f and \ref{fig::10}i). The main difference between the pressure and the temperature field is that the former is fairly uniform inside  the bubble, while the latter is a function of space. The temperature field is continuous across the bubble interface; therefore, a thin thermal boundary layer insures a smooth transition between the inside and the outside of the bubble wall. This boundary layer can be better discerned as $\dot{R}$ decreases (figures \ref{fig::10}f and \ref{fig::10}i) so that the bubble wall motion is isothermal and not adiabatic.  The temperature of the liquid hardly increases, even at the close vicinity of the interface owing to the much higher thermal conductivity of the liquid.
\begin{figure}
	\centering\includegraphics[width=\textwidth]{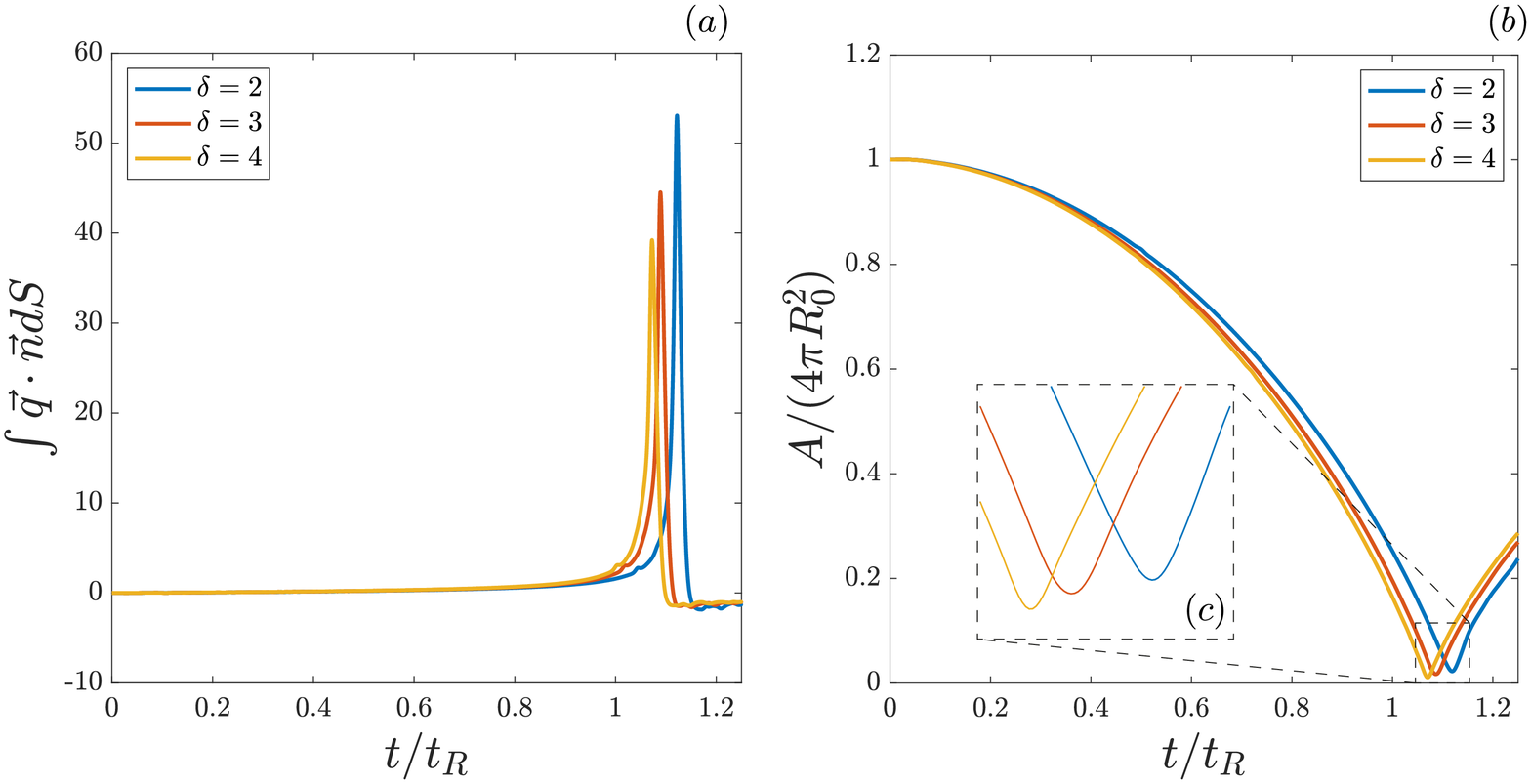}
	\caption{(a) Dimensionless heat flux across the bubble surface with respect to time, calculated using equation \ref{heatflux}, for the different stand-off ratios. (b) Normalised surface area of the bubble with respect to time, for the different values of the stand-off ratio. (c) Zoom around the moment of bubble collapse, showing the minimum surface area achieved for different stand-off ratios.}
	\label{fig::11}
\end{figure}

It is of interest to check the heat flux across the bubble interface, for the different stand-off ratios, and check whether the difference in the bubble shape, and thus surface area, affects it or not. Let the bubble be our control volume. In the absence of mass transfer effects, it is considered as a closed thermodynamic system. The first law of thermodynamics is therefore written as,
\begin{equation}
	dU = \delta Q - \delta W,
	\label{1stlaw}
\end{equation}
where $U=\rho e$ is the internal energy of the bubble, $Q$ the heat supplied to the system and $W$ the mechanical work done by the system due to pressure differences. Derived with respect to time, each of the above terms is expressed as follows,
\begin{align}
	\frac{dU}{dt} &= \frac{d}{dt}\iiint\rho edV,\\
	\frac{\p Q}{\p t} &= -\oiint \boldsymbol{q}\cdot\boldsymbol{n}dS,\\
	\frac{\p W}{\p t} &= p\frac{dV}{dt}.
\end{align}
For a perfect gas, $\rho e = p/\left(\gamma - 1\right)$, and since the pressure inside the bubble is uniform, equation \ref{1stlaw}, derived with respect to time, yields an expression for the heat flux across the bubble interface,
\begin{equation}
	-\oiint \boldsymbol{q}\cdot\boldsymbol{n}dS = \frac{1}{\gamma - 1}V\frac{dp}{dt} + \frac{\gamma}{\gamma - 1}p\frac{dV}{dt}.
	\label{heatflux}
\end{equation}

Figure \ref{fig::11}a shows the heat flux across the bubble surface for the different stand-off ratios $\delta$. As previously mentioned, $\delta$ is the only parameter that changes between the simulations, therefore the heat flux appears to be a function of it. Namely, the heat flux seems to increase with decreasing $\delta$. Figure \ref{fig::11}b shows the normalised surface area of the bubble with respect to time. As the bubble collapses, it deviates from the spherical shape due to the previously discussed mechanisms. This deformation is more prominent for smaller $\delta$, translated by a bigger bubble surface area as can be seen from the inset (figure \ref{fig::10}c). This leads to further contact between both phases, and therefore to a larger heat flux.

\section{Conclusions and outlook}\label{sec::6}
In this paper, we presented a generalisation of the all-Mach solver \cite{jcp-fuster-2018}, previously adiabatic, so that it takes into account heat diffusion between the different phases. Therefore, we derived a two-way coupled system of equations for pressure and temperature, which was then solved implicitly using a multigrid solver. Different test cases were proposed to validate the correct implementation of the thermal effects. An Epstein-Plesset like problem is studied, where a temperature gradient exists across the bubble wall and drives the flow. The temporal evolution of the bubble radius is shown to compare well with a spectral method solution. The code also reproduces free small amplitude oscillations of a spherical bubble. As the equilibrium radius decreases, the Peclet number associated with the bubble oscillations also decreases. Analytical solutions therefore predict a transition from adiabatic to isothermal oscillations. A good agreement between the simulations and the theory is achieved. In addition, we show results of a single sonoluminescent bubble (SBSL) in standing waves, where the result of the DNS is compared with that of other methods in the literature. Besides capturing the thermal effects, our code exhibits the strongest damping as compared to the other tested methods because it solves for the compressible effects in the liquid, and thus accurately predicts the emission of shock waves. Using the solver, the spherical Rayleigh collapse was studied for a wide range of pressure ratios showing a comparison between thermal and adiabatic simulations. For high pressure ratios, the collapses became much more violent and tended to the adiabatic limit, even when thermal diffusion was taken into account. Finally, the axisymmetric collapse of a bubble near a rigid boundary was studied, giving the change of heat flux as a function of the stand-off distance.

The present work extends the applicability of the all-Mach solver to the simulation of compressible multiphase flows where thermal effects are relevant. Applications could be the study of thermal ablation by means of bubble inception and collapse close to boundaries. Also, the current implementation is one step further towards the simulation of boiling flows for example. Future work could be the extension of the all-Mach solver to include mass transfer and phase change, which would allow the simulation of a plethora of physical flows involving all the previous effects.

\section*{Acknowledgments}
The authors would like to thank Andrea Prosperetti for fruitful discussions. The research leading to these results has received funding from the European Union’s Horizon 2020 Research and Innovation programme under theMarie Skłodowska-Curie Grant Agreement No 813766. This work was carried out on the national e-infrastructure of SURFsara, a subsidiary of SURF cooperation, the collaborative ICT organization for Dutch education and research.

\section*{Code availability}
All the codes necessary to reproduce the results presented in this article are available in {\fontfamily{qzc}\selectfont Basilisk }'s sandbox \cite{ysaade}.

\section*{Author ORCIDs}\noindent
Youssef Saade, \href{https://orcid.org/0000-0002-3760-3679}{https://orcid.org/0000-0002-3760-3679};\\
Detlef Lohse, \href{https://orcid.org/0000-0003-4138-2255}{https://orcid.org/0000-0003-4138-2255};\\
Daniel Fuster, \href{https://orcid.org/0000-0002-1718-7890}{https://orcid.org/0000-0002-1718-7890}.

\appendix
\section{Spectral method}\label{appA}
With the assumptions we made in section \ref{sec::4.1}, equations \ref{enthalpy}--\ref{radiusT} are reduced to the following, assuming spherical symmetry,
\begin{equation}
	\frac{\gamma}{\gamma - 1}\frac{p}{T}\left[\frac{\p T}{\p t} + \frac{\gamma - 1}{\gamma p}k_g\left(\frac{\p T}{\p r}\right)^2\right] = \frac{k_g}{r^2}\frac{\p}{\p r}\left(r^2\frac{\p T}{\p r}\right),
\end{equation}
\begin{equation}
	\dot{R} = \frac{\gamma - 1}{\gamma p}\kappa_g\frac{\p T}{\p r}\biggr\rvert_R.
\end{equation}
It is numerically convenient to treat the problem as that of a fixed boundary, so we use the following coordinate mapping,
\begin{equation}
	y = \frac{r}{R(t)}.
\end{equation}
The equations are then transformed to,
\begin{equation}
	\frac{\gamma}{\gamma - 1}\frac{p}{T}\left[\frac{\p T}{\p t} + \frac{\gamma - 1}{\gamma p}\frac{\kappa_g}{R^2}\left(\frac{\p T}{\p y}\right)^2 - \frac{\dot{R}}{R}y\frac{\p T}{\p y}\right] = \frac{\kappa_g}{R^2y^2}\frac{\p}{\p y}\left(y^2\frac{\p T}{\p y}\right),
	\label{eq::a4}
\end{equation}
\begin{equation}
	\dot{R} = \frac{\gamma - 1}{\gamma p}\frac{\kappa_g}{R}\frac{\p T}{\p y}\biggr\rvert_{y = 1}.
	\label{eq::a5}
\end{equation}
We now attempt to solve the advection-diffusion problem, taking into account the change of boundary via a spectral method.
To that end, we expand the temperature into Chebyshev polynomials,
\begin{equation}
	T(t,y) = \sum_{n=0}^{N}a_n(t)T_{2n}(y),
	\label{cheby}
\end{equation}
where $T_{2n}$ are the Chebyshev polynomials. Notice that only even polynomials are used so as to enforce the symmetry boundary condition at $y=0$. We substitute expansion \ref{cheby} into equation \ref{eq::a4} and the result is evaluated at the Gauss-Lobatto collocation points $y_n$,
\begin{equation}
	y_n = \cos\left(\frac{n\pi}{2N}\right), \ \ n = 1, 2...,N.
\end{equation}
This yields $N$ coupled ODEs of the type $\dot{a}_n=f(a_n)$, for the $N+1$ coefficients. The last equation is constructed from the constant temperature boundary condition at $y = 1$, which we derive with respect to time,
\begin{equation}
	\sum_{n=0}^{N}\dot{a}_nT_{2n}(1) = 0.
\end{equation}
With the added ODE for the bubble radius, equation \ref{eq::a5}, we have a system of N+2 coupled first order ODEs, which is linearised and then implicitly integrated in time.

\section{Standing wave in pure liquid}\label{appB}
\begin{figure}
	\centering\includegraphics[width=0.6\textwidth]{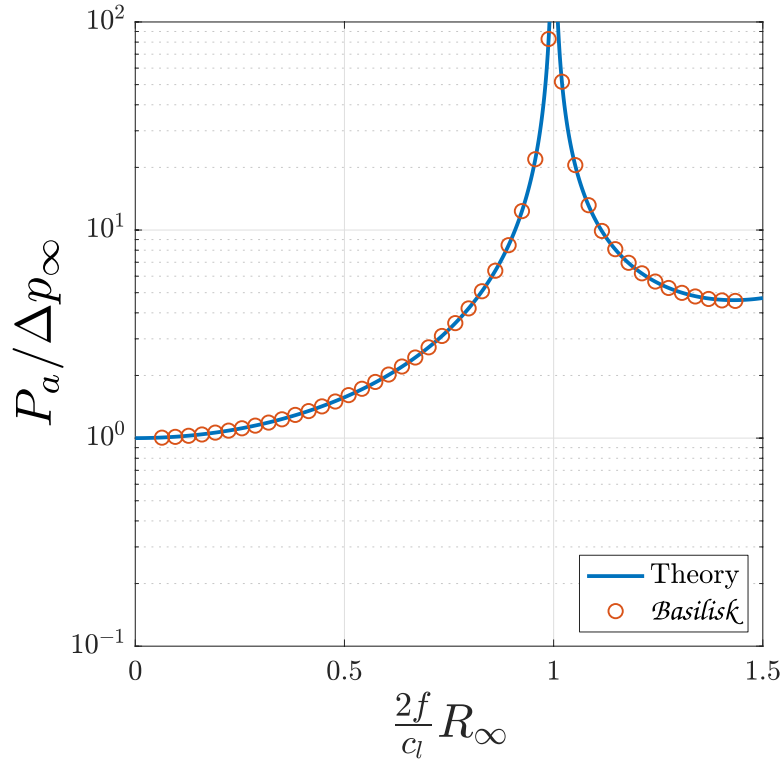}
	\caption{The amplitude of the standing pressure wave $\Delta p_{l,0}$ at $r = 0$, normalised by the amplitude of the driving signal $\Delta p_\infty$, as a function of the dimensionless parameter $2fR_\infty/c_l$. The solid line is computed using equation \ref{swave}.}
	\label{fig::6}
\end{figure}
The inviscid Euler equations, in a spherically symmetric framework, are written as
\begin{equation}
	\frac{1}{\rho_l c_l^2}\frac{\p p_l}{\p t} = -\frac{1}{r^2}\frac{\left(u_lr^2\right)}{\p r},
\end{equation}
\begin{equation}
	\rho_l\frac{\p u_l}{\p t} = -\frac{\p p_l}{\p r}.
\end{equation}
The boundary conditions are
\begin{align}
	u_l(r = 0, t) &= 0 ,\\
	p_l(r = R_\infty,t) &= p_{l,\infty} + \Delta p_\infty\sin\left(\omega t\right),\label{dirichlet}
\end{align}
where $p_\infty$ is the atmospheric pressure, $\Delta p_\infty$ the driving amplitude and $\omega$ the frequency of the acoustic signal. The standing wave solution is \cite{jasa-fuster-2011,theoac},
\begin{align}
	u_l(r,t) &= \frac{\Delta p_\infty}{\rho_l c_l}\frac{1}{\sin \left(kR_\infty\right)}\frac{R_\infty}{r}\left[\cos\left(kr\right) - \frac{\sin\left(kr\right)}{kr}\right]\times\cos\left(\omega t\right),\label{swavesol1}\\
	p_l(r,t) &= p_{l,\infty} + \Delta p_\infty\frac{1}{\sin \left(kR_\infty\right)}\frac{R_\infty}{r}\sin\left(kr\right)\times\sin\left(\omega t\right)\label{swavesol2},
\end{align}
where $k = \omega/c_l$ is the wavenumber. The amplitude of the standing pressure wave at the centre of the flask thus is
\begin{equation}
	\frac{p_l(r = 0) - p_{l,\infty}}{\Delta p_\infty} = \frac{P_a}{\Delta p_\infty} = \frac{kR_\infty}{\sin \left(kR_\infty\right)}.
	\label{swave}
\end{equation}
We test the capability of our code to produce the correct amplitude of a standing pressure wave at $r = 0$. The frequency of the driving signal is set to $f = 10000 \ Hz$, and its amplitude to $\Delta p_\infty = 1 \ atm$. Equation \ref{dirichlet} is set as a Dirichlet boundary condition at $r = R_\infty$. Simulations are performed for multiple values of $R_\infty$. Figure \ref{fig::6} shows a perfect agreement between equation \ref{swave} and our code. The theory predicts resonance for $2fR_\infty/c_l = 1$ which is also observed in our simulations.

\bibliography{mybibfile}

\end{document}